\documentclass{article}
\usepackage{graphicx}

\usepackage{hyperref}

\usepackage[
backend=biber,
style=phys,
biblabel=brackets,
citestyle=numeric-comp,
sorting=none,
minnames=4,
maxnames=4,
]{biblatex}

\DeclareFieldFormat{titlecase}{#1}

\usepackage[T1]{fontenc}
\usepackage{amsmath}  
\usepackage{amssymb}
\usepackage{comment}
\usepackage{color}

\begin{document}
\title{Fermiology, charge transfer energy, and robust paramagnons in high-$T_c$ cuprate superconductors}
\author{\normalsize{Maciej Fidrysiak}\\\normalsize{Institute of Theoretical Physics, Jagiellonian University,}\\\normalsize{ul. Łojasiewicza 11, 30-348 Krak\'ow, Poland}}
\date{}

\maketitle

\begin{abstract}
  Copper-oxide high-temperature (high-$T_c$) superconductors host robust paramagnon excitations whose propagation energies are insensitive to hole concentration and correlate with maximal measured superconducting transition temperatures. Given variation of electronic structure across (and within) cuprate families, elucidation of the relationship between microscopic parameters relevant to high-$T_c$ superconductivity and paramagnon dynamics remains a key challenge to theory. Employing canonical Hubbard- and $t$-$J$-$U$ models of a $\mathrm{CuO_2}$ plane,  we relate robust paramagnon energies to high-$T_c$ fermiology (via the ratio $r \equiv t^\prime/|t|$ of next-nearest- to nearest-neighbor hopping integrals) and charge transfer energy, $\Delta_\mathrm{CT}$. It is shown that variation of $r$ and $\Delta_\mathrm{CT}$ between materials has an opposite effect on paramagnon energy, rationalizing comparable bandwidth of magnetic excitations across multiple cuprates. Utilizing empirical values of $r$ and $\Delta_\mathrm{CT}$ as input to theory, we address magnetic dynamics in Bi-cuprate family representatives with up to three $\mathrm{CuO_2}$ planes, and demonstrate quantitative (within $6\,\%$ margin) agreement of calculated paramagnon energies with experiment. Our work offers a route toward quantitative control of robust paramagnon physics in strongly-correlated electron systems.
\end{abstract}

\section{Introduction}

High-temperature superconductivity (high-$T_c$ SC) in layered copper oxides is induced by introduction of either holes or electrons into antiferromagnetic (AF) parent compounds \cite{Spalek2022}. Close proximity of SC to AF state motivates theoretical investigation of magnetic excitations across high-$T_c$ phase diagram, and points toward their plausible relevance to high-$T_c$ phenomenology. Upon hole doping, the low-energy collective modes of parent AF insulators (magnons) evolve into collective excitations of the paramagnetic state (paramagnons). Whereas magnons emerge as long-lived modes with well defined energies, paramagnons share phase space with incoherent continuum excitations and are prone to overdamping due to kinematically allowed decays into particle-hole pairs. There is, however, extensive experimental evidence (cf.~\cite{Dean2013} and subsequent work) that the latter scenario is not realized in the cuprates, and paramagnon propagation energies measured along the anti-nodal Brillouin-zone direction remain insensitive to hole concentration, from underdoped to overdoped regime. This circumstance, referred to as robust paramagnon behavior, has been recently linked to high-$T_c$ SC by demonstration of empirical correlation between paramagnon energies and maximal SC transition temperatures in multiple families of cuprates \cite{Wang2022}. Due to limited theoretical insights into the factors governing robust paramagnon dynamics, the microscopic origin of this relationship remains unclear. Whereas AF magnons are well described in terms of spin-only Heisenberg-type Hamiltonians with known set of exchange integrals \cite{Coldea2001}, microscopic modeling of paramagnons necessarily involves strongly-correlated itinerant electrons. This renders magnetic excitations sensitive not only to bare magnitude of the AF exchange, but also to the underlying electronic structure that varies significantly across (and within) high-$T_c$ cuprate families. A quantitative characterization of single-particle electronic properties is provided by angle-resolved photoemission spectroscopy (ARPES) that maps both Fermi-surface geometry and dispersion of Fermi quasiparticles. In terms of effective tight-binding (TB) parametrization of ARPES spectra, the ratio $r \equiv t^\prime/|t|$ of next-nearest to nearest-neighbor hopping integrals falls into a broad range $\sim 0.1$-$0.45$, depending on the cuprate and TB model details \cite{Yoshida2006,Vishik2014,Markiewicz2005,Lee2006,Zhong2018}. TB analysis of density-functional theory (DFT) band structures yields consistent results \cite{Markiewicz2005,Das2012,Pavarini2001,Photopoulos2019}. The range parameter, $r$, provides information about renormalized Fermi quasiparticles and is thus, in essence, a single-particle characteristic. Since high-$T_c$ cuprates fall into charge-transfer regime of the Zaanen-Sawatsky-Allen (ZSA) classification \cite{Zaanen1985}, another quantity relevant for paramagnon dynamics is charge transfer energy, $\Delta_\mathrm{CT}$, here understood as a distance between centers of the Zhang-Rice-singlet- and upper-Hubbard bands. This is because, within one-band mapping of the underlying three-band Hubbard model of a $\mathrm{CuO_2}$ plane, $\Delta_\mathrm{CT}$ governs effective on-site electron-electron interaction, $U$ \cite{Feiner1996}. The latter relationship allows us to estimate microscopic interaction parameters for concrete materials, supplementing the single-particle properties determined by ARPES. Besides differences in charge-transfer energy between cuprate families \cite{Ruan2016}, scanning transmission electron microscopy coupled with electron energy-loss spectrometry (STEM-EELS) \cite{Wang2023} reveals also systematic variation of $\Delta_\mathrm{CT}$ among representatives of the same family, calling for an investigation of its relation to paramagnon dynamics. 

\begin{figure}[!ht]
\centerline{%
  \includegraphics[width=0.7\linewidth]{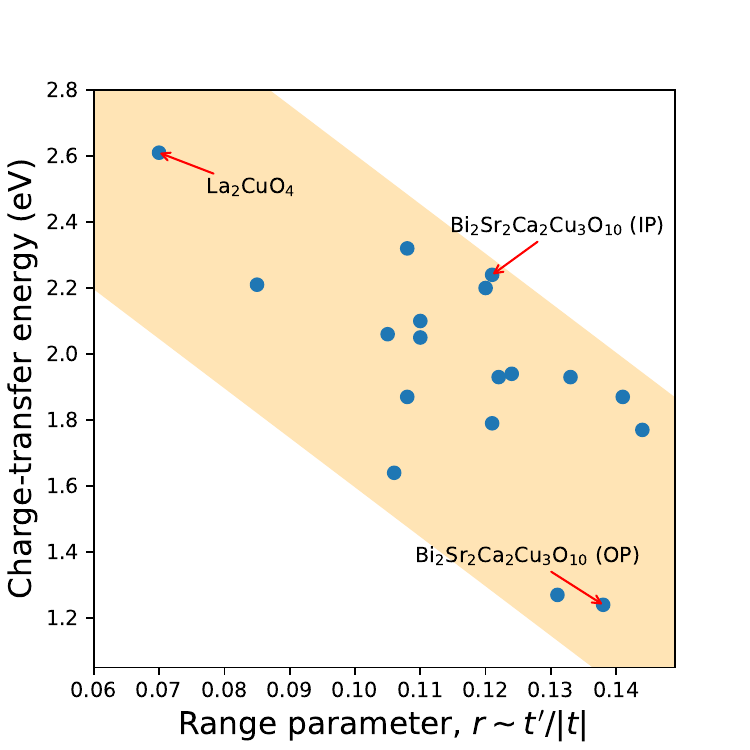}}
\label{fig:r_vs_ctgap}
\caption{Illustration of the negative correlation between range parameter, $r \sim t^\prime/|t|$, and charge-transfer energy in high-$T_c$ cuprates, composed based on data of Ref.~\cite{Weber2012}. Symbols represent first-principle calculation results for multiple cuprates and shaded orange region is guide to the eye. The named copper oxides, La$_2$CuO$_4$ (LSCO) and Bi$_2$Sr$_2$Ca$_2$Cu$_3$O$_{10}$ (Bi2223), are positioned on the opposite ends of $r$ vs. charge-transfer energy phase diagram. Here IP and OP refer to inner- and outer planes, respectively. Those two compounds have the same paramagnon bandwidth within error bars, $311(4)\,\mathrm{meV}$ (LSCO) and $324(15)\,\mathrm{meV}$ (Bi2223) \cite{Wang2022}, pointing toward an intricate relationship between electronic structure and robust paramagnon dynamics.}
\label{fig:tb_mapping}
\end{figure}

In this work we aim to reconcile universal persistence of paramagnons in hole-doped cuprates with known variation of their electronic structure, and establish a microscopic relationship between fermiology, charge transfer energy, and energies of magnetic excitations. Canonical one-band Hubbard- and $t$-$J$-$U$ models are employed and analyzed using variational wave function (VWF) approach, combined with the expansion in inverse number of fermionic flavors, $1/\mathcal{N}_f$. The latter, VWF+$1/\mathcal{N}_f$ method, was introduced and tested elsewhere \cite{Fidrysiak2021}. We first construct a mapping between two common TB parameterizations of high-$T_c$ fermiology (incorporating hopping integrals up to either two or three nearest-neighbor sites) by matching their Fermi surfaces. This is a prerequisite for a quantitative discussion since the magnitude of the range parameter $r$ is highly sensitive to the choice of parameterization. Subsequently, incorporating strong electronic correlations, we analyze paramagnon dynamics on both hole- and electron-doped side of high-$T_c$ phase diagram. We find that only the more sophisticated (three-parameter) TB parameterization properly accounts for paramagnon persistence in entire high-$T_c$ regime, pointing toward relevance of longer-range hopping integrals. Furthermore, we demonstrate that increase of either $r$ or $\Delta_\mathrm{CT}$ results in softening of magnetic excitations. Taking into account numerical evidence for a negative correlation between these two quantities \cite{Weber2012}, illustrated in Fig.~\ref{fig:r_vs_ctgap}, our result indicates that the contributions to paramagnon energy from variation of $r$ and $\Delta_\mathrm{CT}$ have opposite signs and tend to cancel each other. This rationalizes comparable characteristic magnetic energy scale across materials with substantially distinct electronic structure, observed in experiment. Indeed, six out of twelve cuprates considered in Ref.~\cite{Wang2022} host magnons/paramagnons with remarkably similar maximal energies, falling in the range $290$-$324\,\mathrm{meV}$. Among them, La$_2$CuO$_4$ and  Bi$_2$Sr$_2$Ca$_2$Cu$_3$O$_{10+\delta}$ have the same magnetic bandwidth within error bars, $311(4)\,\mathrm{meV}$ and $324(15)\,\mathrm{meV}$ \cite{Wang2022}, while being positioned on the opposite ends of the high-$T_c$ cuprate $r$ vs. $\Delta_\mathrm{CT}$ phase diagram, cf. Fig.~\ref{fig:r_vs_ctgap}. Finally, we carry out a unified analysis of paramagnon dynamics for the first three ($n=1, 2, 3$) representatives of Bi$_2$Sr$_2$Ca$_{n-1}$Cu$_n$O$_{2n+4+\delta}$ series (Bi2201, Bi2212, and Bi2223, respectively). Utilizing empirical values of $r$ and $\Delta_\mathrm{CT}$ as input to theory, we calculate robust paramagnon energies and compare them with available resonant inelastic $x$-ray scattering (RIXS) data. Our results agree quantitatively with measured values within $6\,\%$ margin, relating properties obtained by three distinct experimental probes, and providing insight into systematic evolution of magnetic excitations within Bi-family of cuprates.

\section{Tight-binding parameterizations of high-$T_c$ superconductors}
\label{sec:tb_parametrization}

A general TB parametrization of high-$T_c$ fermiology is based on the Hamiltonian $\hat{T} \equiv \sum_{i \neq j, \sigma} t_{ij} \hat{a}_{i\sigma}^\dagger \hat{a}_{j\sigma}$, where $\hat{a}_{i\sigma}^\dagger$ ($\hat{a}_{i\sigma}$) are spin-$\sigma$ fermionic creation (annihilation) operators for site $i$ of a square lattice, and $t_{ij}$ denote hopping integrals. In theoretical work, particularly that based on state-of-the art numerical methods, it is common to retain only nearest- and next-nearest-neighbor integrals in $\hat{T}$. We refer to such a simplification as two-parameter (2P) model, and denote the nonzero hopping integrals as $t_\mathrm{2P}$ and $t^\prime_\mathrm{2P}$, respectively. A more realistic, three-parameter (3P) model, includes hopping integrals up to second-nearest neighbor sites, i.e. $t_\mathrm{3P}$, $t^\prime_\mathrm{3P}$, and $t^{\prime\prime}_\mathrm{3P}$. Hereafter we adopt leading-order theoretical relation $t^{\prime\prime}_\mathrm{3P} = - 0.5 \cdot t^{\prime}_\mathrm{3P}$ \cite{Pavarini2001}.

As we elaborate below, the magnitude of longer-range hopping integrals, controlled the range parameter $r_\mathrm{\alpha P} \equiv t^\prime_\mathrm{\alpha P}/|t_\mathrm{\alpha P}|$ (with $\alpha = 2$ or $3$, depending on parametrization), is among key factors affecting paramagnon energies. It is thus noteworthy that 2P and 3P TB models, employed for modeling the same reference data, yield substantially different values of $r_\mathrm{\alpha P}$. We demonstrate this explicitly by constructing a mapping between the two parametrizations. This is carried out by defining 2P and 3P energy dispersion relations, $\epsilon^\mathrm{2P}_\mathbf{k} = t_\mathrm{2P} f_1(\mathbf{k}) +  t_\mathrm{2P}^\prime f_2(\mathbf{k}) - \mu_\mathrm{2P}$ and $\epsilon^\mathrm{3P}_\mathbf{k} = t_\mathrm{3P} f_1(\mathbf{k}) +  t_\mathrm{3P}^\prime f_2(\mathbf{k})  -\frac{1}{2}  t_\mathrm{3P}^{\prime} f_3(\mathbf{k}) - \mu_\mathrm{3P}$, respectively, with $\mathbf{k} = (k_x, k_y)$ being in-plane wave vector. We have introduced a notation $f_1(\mathbf{k}) \equiv 2 \cdot (\cos k_x + \cos k_y)$, $f_2(\mathbf{k}) \equiv 4 \cos k_x \cdot \cos k_y$, and $f_3(\mathbf{k}) \equiv 2 \cdot (\cos 2 k_x + \cos 2 k_y)$, with lattice spacing set to unity. The 2P and 3P chemical potentials, $\mu_\mathrm{2P}$ and $\mu_\mathrm{3P}$, are determined by the condition that total electronic density $n_e$ (here understood as the number of carriers per Cu site) is fixed and equal for both models. In effect, $\mu_\mathrm{2P} = \mu_\mathrm{2P}(t_\mathrm{2P}, t^\prime_\mathrm{2P}, n_e)$ and $\mu_\mathrm{2P} = \mu_\mathrm{3P}(t_\mathrm{3P}, t^\prime_\mathrm{3P}, n_e)$ become implicit functions of the respective hopping integrals and density. Our mapping of 3P onto 2P model is based on the cost function

\begin{align}
  S = \frac{1}{N} \sum_\mathbf{k} \theta( - \epsilon^{2\mathrm{P}}_\mathbf{k} \cdot \epsilon^{3\mathrm{P}}_\mathbf{k} ),
  \label{eq:fs_function}
\end{align}

\noindent
where $\theta$ denotes Heaviside step function and $N = 400 \times 400$ is the square lattice size. Physically, Eq.~(\ref{eq:fs_function}) measures mismatch between 2P and 3P Fermi surfaces via fraction of single-particle states that are occupied within one model, but empty within the other. Note that $S$ is insensitive to rescaling $t_\mathrm{\alpha P} \rightarrow c_{\alpha P} \cdot t_\mathrm{\alpha P}$ and $t^\prime_\mathrm{\alpha P} \rightarrow c_\mathrm{\alpha P} \cdot t^\prime_\mathrm{\alpha P}$ by coefficients $c_{\alpha\mathrm{P}} > 0$, selected independently for $\alpha = 2$ and $3$. This becomes apparent after noting that $\mu_\mathrm{\alpha P}(c_\mathrm{\alpha P} \cdot t_\mathrm{\alpha P}, c_\mathrm{\alpha P} \cdot t^\prime_\mathrm{\alpha P}, n_e) = c_{\alpha P} \cdot \mu_\mathrm{\alpha P}(t_\mathrm{\alpha P}, t^\prime_\mathrm{\alpha P}, n_e)$ so that dispersion relations $\epsilon^\mathrm{\alpha P}_\mathbf{k}$ are homogeneous functions of $c_\mathrm{\alpha P}$. In effect, assuming $t_{\alpha\mathrm{P} } < 0$, the right-hand-side of Eq.~(\ref{eq:fs_function}) depends only on  ratios $r_\mathrm{2\mathrm{P} }$  and $r_\mathrm{3\mathrm{P}}$, as well on density $n_e$, i.e. $S = S(r_\mathrm{2P}, r_\mathrm{3P}, n_e)$. This circumstance allows us to set $t_\mathrm{2P} = t_\mathrm{3P} \equiv t$ in the subsequent analysis, without affecting the TB model mapping. For technical reasons, chemical potentials are evaluated at low finite temperature, $k_B T = 10^{-6} |t|$.

\begin{figure}[!ht]
\centerline{%
\includegraphics[width=\linewidth]{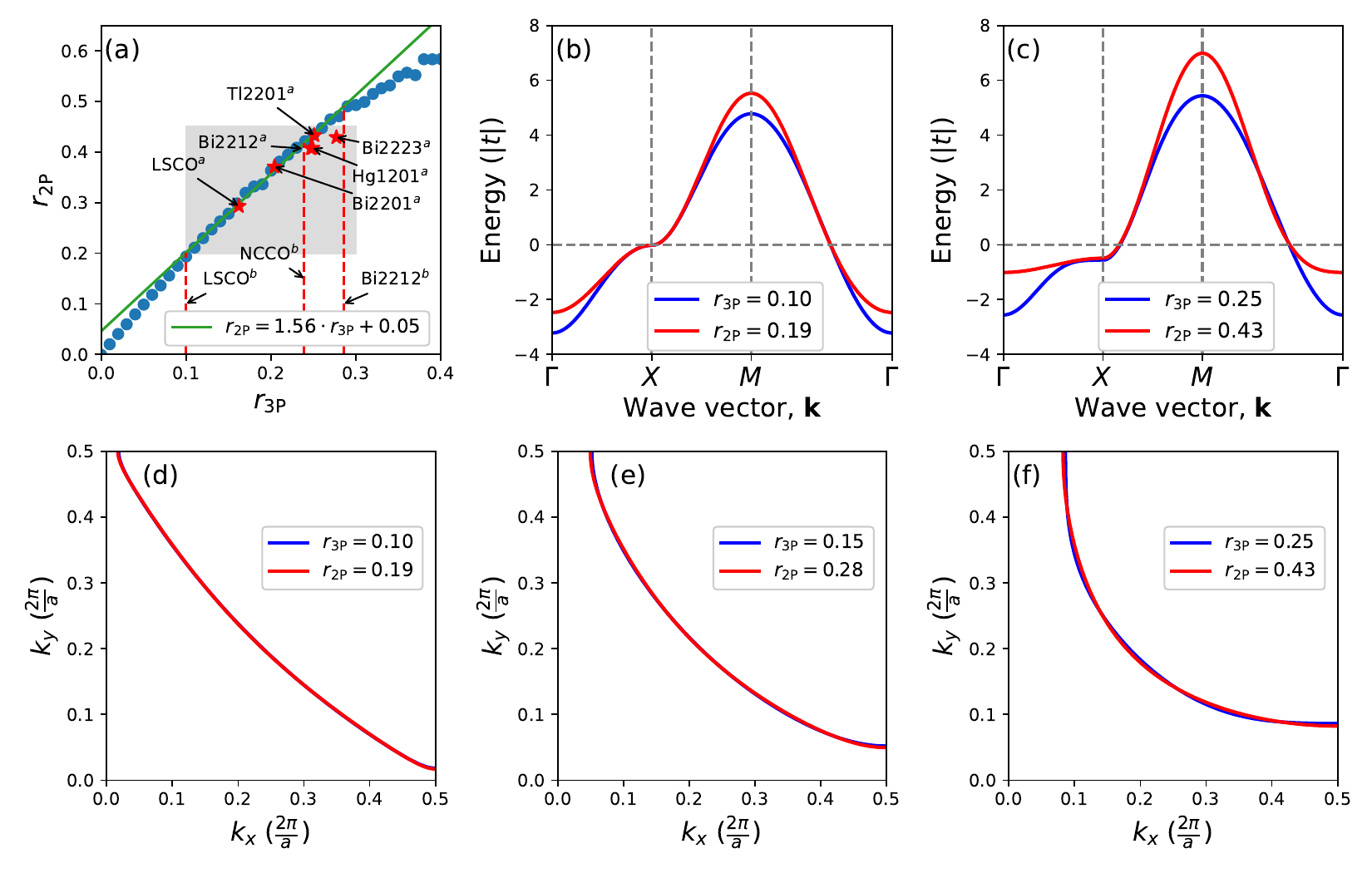}}
\caption{Comparison between the two-parameter (2P) and three-parameter (3P) tight-binding models of high-$T_c$ cuprates for representative density $n_e = 0.85$. Solid circles in panel (a) show the established correspondence between range parameters $r_\mathrm{2P}$ and $r_\mathrm{3P}$. Solid line represents linear fit of those data for $r_\mathrm{3P} \in [0.1, 0.3]$. Red stars are experimental ARPES data for multiple cuprates (superscript ``\emph{a}'') \cite{Lee2006}, whereas vertical lines mark $t^\prime/|t|$ estimated using a more general 3P-model-like parameterization (superscript ``\emph{b}'') \cite{Markiewicz2005}. The high-$T_c$ regime, enclosing available experimental data, is indicated by grey area. The comparison between band structures [panels (b)-(c)] and Fermi surfaces [panels (d)-(f)] of both models for values of $r_\mathrm{2P}$ and $r_\mathrm{3P}$ related via mapping of panel (a) is also shown (employed parameters are detailed inside the figure; displayed $r_\mathrm{2P}$ values have been truncated to two significant digits). In-plane wave vector components in panels (d)-(f) are given in the units of $2\pi/a$ with $a$ being square lattice spacing.}
\label{fig:tb_mapping}
\end{figure}

Figure~\ref{fig:tb_mapping}(a) details the correspondence between 2P and 3P models (blue symbols), established by minimization of $S$ [cf. Eq.~(\ref{eq:fs_function})] over $r_\mathrm{2P}$ for given $r_\mathrm{3P}$. Electronic density is set to a representative value $n_e = 0.85$ and fixed throughout the optimization procedure. Relationship $r_\mathrm{2P} \approx 1.56 \cdot r_\mathrm{3P} + 0.05$ (green line), resulting from linear regression of obtained data in the high-$T_c$ range ($r_\mathrm{3P} = 0.1$-$0.3$), shows that the 2P model yields relative values of next-nearest-neighbor hopping integral systematically larger than the corresponding 3P model. Our result is consistent with former 2P and 3P model analysis of ARPES data for multiple cuprates (red stars) \cite{Lee2006}. By vertical dashed lines, in Fig.~\ref{fig:tb_mapping}(a) we also mark ratios $t^\prime/|t|$, obtained using more general TB models \cite{Markiewicz2005}, comparable to the present 3P scheme. Note that these data cannot be assigned points in $r_\mathrm{3P}$-$r_\mathrm{2P}$ plane, since the corresponding 2P-model values are not available. Figure~\ref{fig:tb_mapping}(a) indicates that the parameter range relevant to most high-$T_c$ cuprates is approximately enclosed by a rectangle within $r_\mathrm{3P}$-$r_\mathrm{2P}$ plane, defined by the conditions $r_\mathrm{3P} \in [0.1, 0.3]$ and $r_\mathrm{2P} \in [0.2, 0.45]$ (cf. grey area).

In the remaining panels of Fig.~\ref{fig:tb_mapping} we carry out a more detailed comparison between the corresponding 2P (red curves) and 3P (blue curves) models for representative values of $r_{2\mathrm{P}}$ and $r_{3\mathrm{P}}$ related via the mapping of panel (a). Their values are provided inside the panels. Figure~\ref{fig:tb_mapping}(b)-(c) shows the band structure along the high-symmetry $\Gamma$-$X$-$M$-$\Gamma$ Brillouin-zone contour. Energies are measured relative to the Fermi level (here set to be zero) and marked by horizontal dashed line. It is apparent that the low-energy quasiparticle energies match closely between the models, and the bandwidths remain similar despite quantitatively distinct dispersions along the $\Gamma$-$M$ and $\Gamma$-$X$ lines. In Fig.~\ref{fig:tb_mapping}(d)-(f), we compare the 2P and 3P model Fermi surfaces within the high-$T_c$ regime of $r_{2\mathrm{P}}$ and $r_{3\mathrm{P}}$, and demonstrate a quantitative agreement between parameterizations.

\section{Microscopic model and method}
\label{sec:model_method}

Resonant paramagnon excitations emerge as a consequence of interplay between itinerant-electron dynamics and electronic correlations. To address those effects, we extend the above 2P and 3P schemes by employing a general $t$-$J$-$U$ Hamiltonian

\begin{align}
  \hat{H}_\mathrm{\alpha P} = \hat{T}_\mathrm{\alpha P} + U \sum_{i} \hat{n}_{i\uparrow} \hat{n}_{i\downarrow} + J \sum_{\langle i, j \rangle} \hat{\mathbf{S}}_i \cdot \hat{\mathbf{S}}_j  - \mu_\mathrm{\alpha P} \sum_{i\sigma} \hat{n}_{i\sigma},
  \label{eq:tju-model}
\end{align}

\noindent
where $\hat{n}_{i\sigma} \equiv \hat{a}_{i\sigma}^\dagger \hat{a}_{i\sigma}$ and $\hat{\mathbf{S}}_i \equiv \frac{1}{2} \sum_{\alpha\beta} \hat{a}_{i\alpha}^\dagger \boldsymbol{\sigma}_{\alpha\beta} \hat{a}_{i\beta} $ denote particle-number- and spin operators, and $\boldsymbol{\sigma} \equiv (\sigma^x, \sigma^y, \sigma^z)$ are Pauli matrices. Summation over nearest-neighbor sites of a square lattice is indicated as $\langle i, j\rangle$. The subscript $\alpha\mathrm{P}$ with $\alpha = 2, 3$ specifies whether two- or three-parameter variant of kinetic energy operator $\hat{T}$ is employed (cf. Sec.~\ref{sec:tb_parametrization}). For compactness of notation, we explicitly incorporate the chemical potential term $\propto \mu_\mathrm{\alpha P}$ into Eq.~(\ref{eq:tju-model}). The rationale behind selection of the $t$-$J$-$U$ Hamiltonian is that it encompasses canonical one-band models of high-$T_c$ SC as limiting cases. For $U \neq 0$ and $J = 0$, it reduces to the Hubbard model, whereas the $t$-$J$ model \cite{Spalek1976,Chao1977} is obtained for $J > 0$ in the $U \rightarrow \infty$ limit. The general $t$-$J$-$U$ Hamiltonian allows a more refined control over electronic correlations and antiferromagnetic exchange than either of its particular limits. Indeed, the leading nontrivial-order canonical perturbation expansion applied to the Hubbard model yields effective magnetic exchange scale $J_\mathrm{eff} = \frac{4 t^2}{U}$ \cite{Spalek1976,Chao1977} that is determined completely by on-site repulsion $U$ and hopping integral $t$. On the other hand, within the $t$-$J$-$U$ model, $J_\mathrm{eff} = \frac{4 t^2}{U} + J$ takes a more elaborate form and may be tuned by means of the explicit exchange $J$, independently of $U$ and $t$ \cite{Fidrysiak2021b}. Moreover, the $t$-$J$-$U$ Hamiltonian has been demonstrated to provide a more accurate description of certain measured properties than either Hubbard- or $t$-$J$ models \cite{Spalek2017}. The concept of effective exchange interaction $J_\mathrm{eff} \equiv \frac{4 t^2}{U} + J$ \cite{Fidrysiak2021b} allows us to carry out a direct comparison between various one-band models relevant to high-$T_c$ SC. Below we focus on the case of Hubbard ($U = 7 |t|$, $J = 0$) and $t$-$J$-$U$ ($U = 15 |t|$, $J = \frac{32}{105} |t| \approx 0.305 |t|$) Hamiltonians, corresponding to the same $J_\mathrm{eff} = \frac{4}{7} |t|$. Assuming a generic value $t = -0.35\,\mathrm{eV}$, this yields $J_\mathrm{eff} = 200\,\mathrm{meV}$.

Magnetic excitations of the model~\eqref{eq:tju-model} are analyzed by means of dynamical spin susceptibility

\begin{align}
  \chi_s(i \omega_n, \mathbf{k}) = \int_0^\beta d\tau \mathrm{e}^{i \omega_n \tau} \frac{1}{N} \sum_{ij} \mathrm{e}^{-i \mathbf{k} (\mathbf{r}_i - \mathbf{r}_j)} \langle \mathcal{T}_\tau \hat{S}^z_i(\tau) \hat{S}^z_j \rangle,
  \label{eq:imagnary-time_susceptibility}
\end{align}

\noindent
where $\beta \equiv (k_B T)^{-1}$ is inverse temperature (in energy units), $k_B$ denotes Boltzmann constant, and $\omega_n = \frac{2 \pi}{\beta} \cdot n$ are bosonic Matsubara frequencies. Imaginary-time dependent spin operators $\hat{S}^z_i(\tau) \equiv \mathrm{e}^{\hat{H}_\mathrm{\alpha P} \cdot \tau} \hat{S}^z_i \mathrm{e}^{-\hat{H}_\mathrm{\alpha P} \cdot \tau}$ inside the thermal expectation value brackets $\langle \ldots \rangle$ are subjected to time ordering by operator $\mathcal{T}_\tau$. Real-frequency dynamical spin susceptibility $\chi_s(\omega, \mathbf{k})$ is obtained from Eq.~(\ref{eq:imagnary-time_susceptibility}) by analytic continuation $i \omega_n \rightarrow \omega + i \epsilon$ with $\epsilon = 0.02 |t|$. To stay clear of ordering instabilities, known to proliferate across the slave-boson \cite{Igoshev2015} and variational-wave-function \cite{Spalek2022} Hubbard-model phase diagram, we carry out subsequent calculations at high temperature $k_B T = 0.35 |t|$. This choice ensures that the $k_B T$ is smaller than the magnetic bandwidth $\sim |t|$, limiting the effects of thermal fluctuations on high-energy paramagnons.

\begin{figure}[!ht]
\centerline{%
\includegraphics[width=12.5cm]{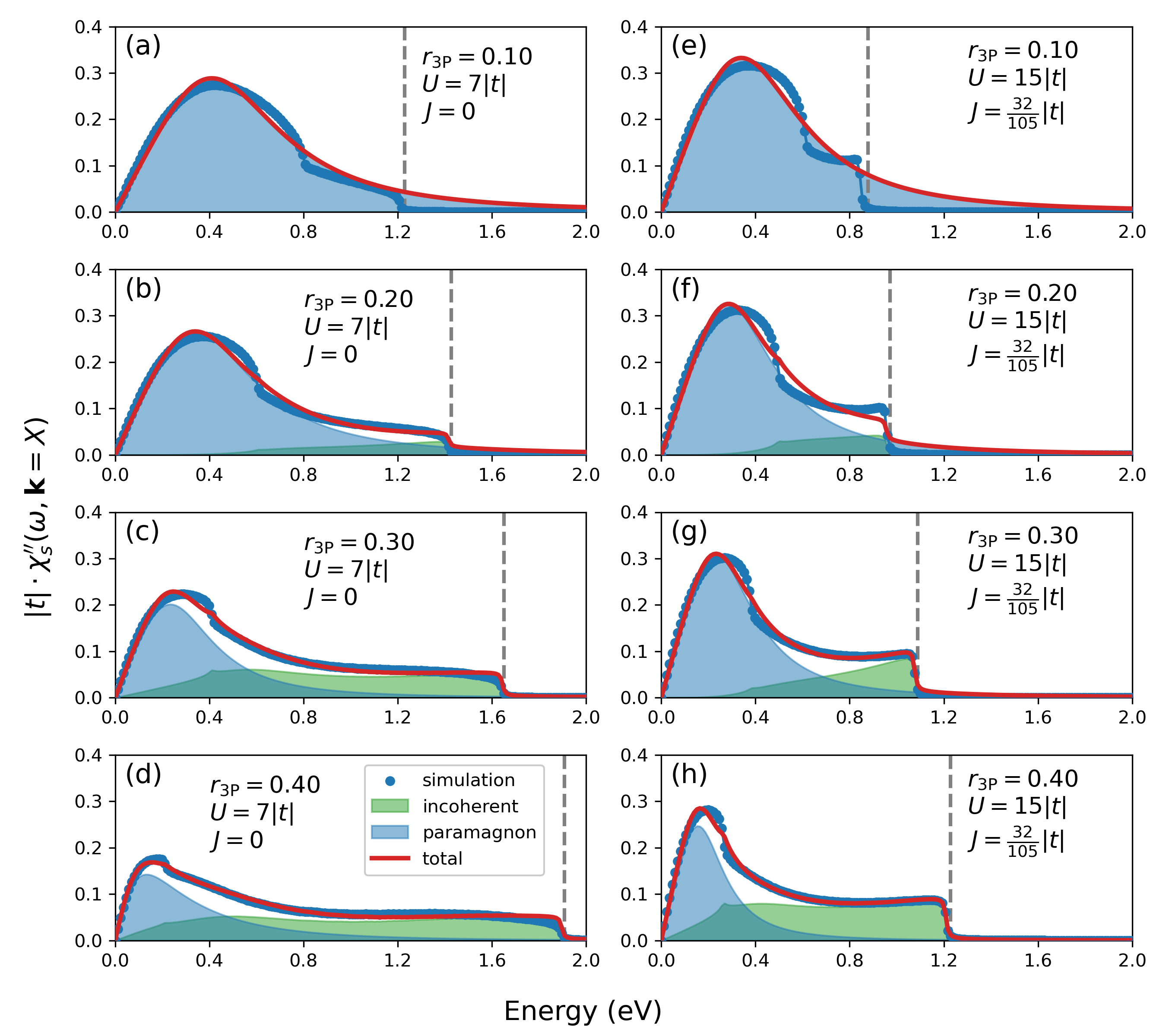}}
\caption{VWF+$1/\mathcal{N}_f$ imaginary part of the dynamical spin susceptibility at the $X$ Brillouin-zone point, obtained for hole-doped 3P model at representative density $n_e = 0.84$. Blue symbols are simulated data, whereas red line represents total model fit in the energy interval $w \in [0, \omega_\mathrm{th}]$. Shaded regions detail the decomposition of the total model signal into paramagnon (blue) and incoherent particle-hole (green) components. The particle-hole continuum excitation threshold, $\omega_\mathrm{th}$, is marked by a vertical dashed line. Panels (a)-(d) correspond to the Hubbard model ($U = 7 |t|$, $J = 0$, and Fermi-surface parameter $r_\mathrm{3P} = 0.1$-$0.4$). Panels (e)-(h) show analogous results for the $t$-$J$-$U$ model ($U = 15 |t|$, $J = \frac{32}{105} |t| \approx 0.305 |t|$, $r_\mathrm{3P} = 0.1$-$0.4$). The parameters are listed inside the figure. 
}
\label{fig:fit}
\end{figure}

Magnetic dynamics of the $t$-$J$-$U$ model~\eqref{eq:tju-model} is analyzed within the framework of VWF+$1/\mathcal{N}_f$ method, combining variational wave function (VWF) scheme with expansion in the inverse number of fermionic flavors ($1/\mathcal{N}_f$). Leveraging the ability of the reference VWF solution to account for strong local correlation effects, $1/\mathcal{N}_f$ expansion around VWF saddle-point state allows us to study the interplay of correlations and long-wavelength fluctuations. VWF+$1/\mathcal{N}_f$ method has been introduced and tested elsewhere \cite{Fidrysiak2021,Spalek2022}, see Appendix~\ref{appendix:method} for outline of its formulation. Figure~\ref{fig:fit} shows VWF+$1/\mathcal{N}_f$ 3P-model imaginary part of dynamical spin susceptibility, $\chi_s(\omega, \mathbf{k} = X)$, obtained for several representative parameter sets (detailed inside the panels) and $t = -0.35\,\mathrm{eV}$. Similar results for the 2P model have been obtained previously (cf., e.g., \cite{Fidrysiak2020}), and we do not include them in Fig.~\ref{fig:fit}. The wave vector is set to $\mathbf{k} \equiv (0.5, 0)$ in reciprocal lattice units (i.e. anti-nodal Brillouin-zone point $X$), corresponding to the high-energy part of the paramagnon spectrum. Panels (a)-(d) show the Hubbard model results ($U = 7 |t|$, $J = 0$) for $r_\mathrm{3P}$ varying in the range $0.1$-$0.4$, as detailed inside the panels. Panels~(e)-(h) present similar analysis for the $t$-$J$-$U$ model ($U = 15 |t|$, $J = \frac{32}{105} |t| \approx 0.305 |t|$).

The VWF+$1/\mathcal{N}_f$ paramagnon spectra (blue symbols) comprise a resonant paramagnon peak, located below $\sim 0.5\,\mathrm{eV}$, and a flat shoulder at larger energies, attributed to incoherent excitations. Depending on the value of $r_\mathrm{3P}$ and $U$, the latter extends up to $\sim 1$-$2\,\mathrm{eV}$ and terminates abruptly at the kinematic threshold for creating particle-hole pairs, $\omega_\mathrm{th}$ (marked by vertical dashed lines). It may be noted that, as a consequence of adopting the same value of effective exchange constant that governs resonant magnetic dynamics, the position of the paramagnon peaks is comparable between the Hubbard- and $t$-$J$-$U$ models. On the other hand, the continuum is controlled by the magnitude of on-site repulsion $U$, rather than $J_\mathrm{eff}$, and occupies a substantially narrower energy region in case of the $t$-$J$-$U$ model (right panels). This is because energy required to excite particle-hole pairs is directly affected by correlation-induced single-particle bandwidth renormalization. The observed qualitatively distinct behavior of low- end high-energy parts of the magnetic spectrum validates our decomposition of simulated data into coherent- and incoherent components. 

To obtain quantitative paramagnon characteristics that allow for a direct comparison with experiment, we carry out a secondary modeling of the $X$-point VWF+$1/\mathcal{N}_f$ data using damped harmonic oscillator (DHO) model

\begin{align}
  \chi^{\prime\prime}_\mathrm{DHO}(\omega, \mathbf{k} = X) = a \cdot \frac{2 \gamma \omega}{\left(\omega^2 - \omega_0^2\right)^2 + 4 \gamma^2 \omega^2},
  \label{eq:DHO_model}
\end{align}

\noindent
where $a$, $\omega_0$, and $\gamma$ denote amplitude, bare paramagnon frequency, and damping coefficient, respectively. Equation~(\ref{eq:DHO_model}) is now universally used for interpretation of empirical magnetic-excitation spectra, superseding formerly employed antisymmetrized Lorentzian function \cite{Lamsal2016}. For $w_0 > \gamma$ the paramagnon is resonant and may be assigned the propagation frequency $\omega_p = \sqrt{\omega_0^2 - \gamma^2}$, whereas for $w_0 < \gamma$ it represents an overdamped excitation with $\omega_p = 0$. The total model of dynamical susceptibility dissipative part reads  $\chi^{\prime\prime}_\mathrm{model} = [\chi^{\prime\prime}_\mathrm{DHO} + \chi^{\prime\prime}_\mathrm{bckg}]$, where $\chi^{\prime\prime}_\mathrm{bckg} = (b + c \cdot \omega^2) \cdot \chi^{\prime\prime}_0$ represents incoherent background, modeled as Lindhard (fermion-loop) susceptibility imaginary part ($\chi^{\prime\prime}_0$) multiplied by a quadratic function redistributing spectral weight between resonant- and incoherent components. Note that only even powers of $\omega$ are retained in the polynomial weight to ensure that $\chi^{\prime\prime}_\mathrm{model}$ remains an odd function of frequency. The nonlinear fit of the VWF+$1/\mathcal{N}_f$ data by this model for $\omega \in [0, \omega_\mathrm{th}]$ is illustrated in Fig.~\ref{fig:fit} by solid red curves. The DHO and incoherent contributions to the total intensity are marked as blue- and green shaded regions, respectively. The model function provides a faithful representation of simulated data in the fitting interval ($0 \leq \omega \leq \omega_\mathrm{th}$) across entire considered range of parameter $r_\mathrm{3P}$. Nonetheless, for $r_\mathrm{3P} \lesssim 0.1$, the DHO model deviates from simulated data at larger energies ($\omega > \omega_\mathrm{th}$). This regime of small $r_\mathrm{3P}$ is, however, relevant only to limited number of materials (including La$_{2-x}$Sr$_x$CuO$_4$, cf. Fig.~\ref{fig:tb_mapping}), for which more realistic models might be necessary to account for the high-energy tail of the magnetic spectral weight. Figure~\ref{fig:fit} also shows that increase of $r_\mathrm{3P}$ results in a systematic enhancement of incoherent particle-hole excitations (green area), as well as in shift of the paramagnon peak (blue area) to lower energies. This effect is quantitatively investigated below for both Hubbard- and $t$-$J$-$U$ models.

\section{Paramagnon dynamics: Hubbard model}
\label{sec:results_Hubbard_model}

\begin{figure}[!ht]
\centerline{%
\includegraphics[width=12.5cm]{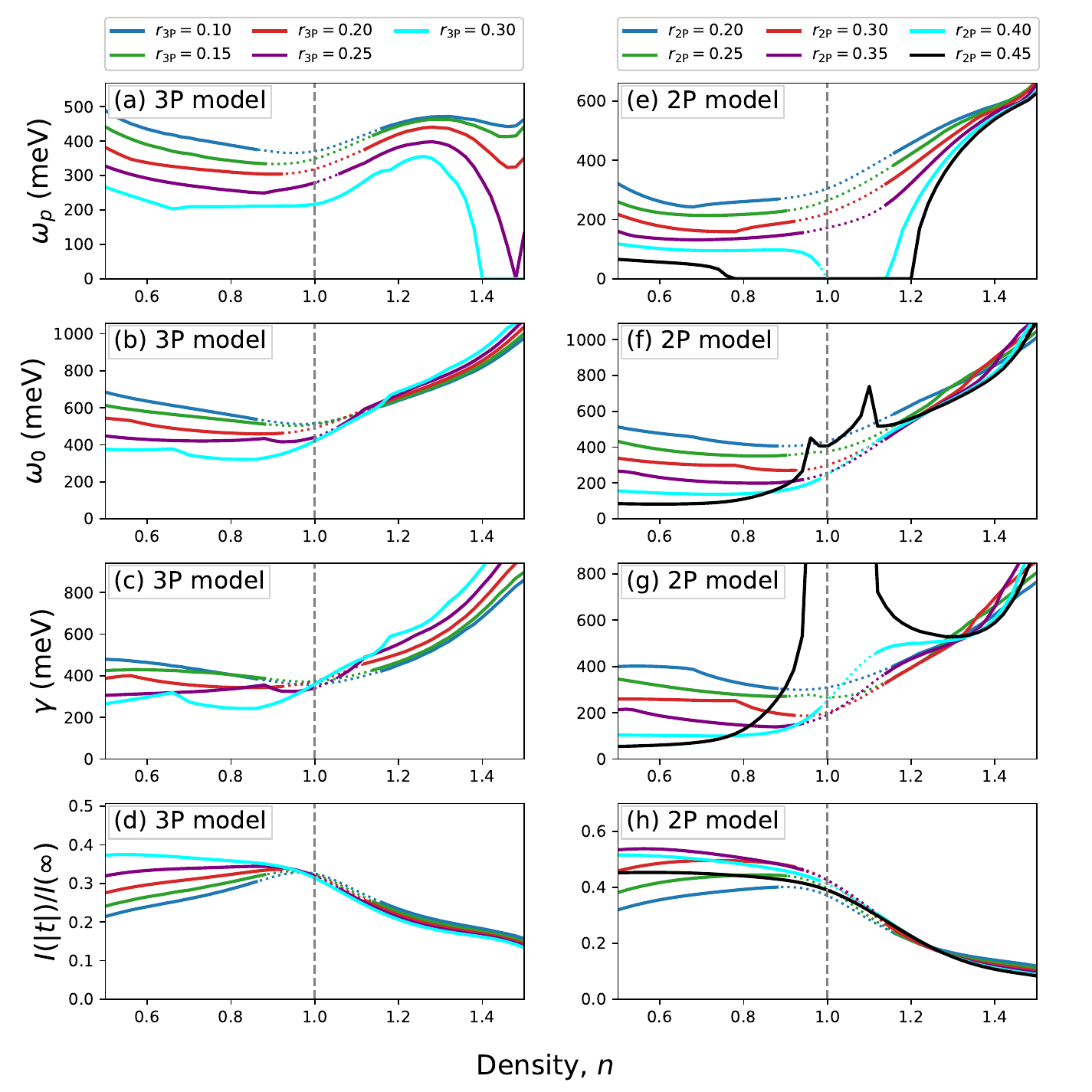}}
\caption{VWF+$1/\mathcal{N}_f$ paramagnon characteristics for the 3P (a)-(d) and 2P (e)-(h) Hubbard model, plotted as a function of electronic density and for parameters $r_\mathrm{2P/3P}$ spanning the range relevant to high-$T_c$ (their values are detailed above the panels). Dotted line segments indicate dynamical instability of the paramagnetic state. The displayed quantities include: paramagnon propagation energy [(a) and (e)], bare energy [(b) and (f)], and damping [(c) and (g)]. Moreover, panels (d) and (h) show and ratio $I(|t|)/I(\infty)$, where $I(\omega) \equiv \int_0^\omega d\nu \chi^{\prime\prime}_s(\nu, X)$ denotes integrated intensity (cf. the text). The simulations have been carried out for $k_B T = 0.35 |t|$, $t = -0.35\,\mathrm{eV}$, and $U = 7|t|$, resulting in effective antiferromagnetic exchange $J_\mathrm{eff} = 200\,\mathrm{meV}$.}
\label{fig:phasediag_Hubbard}
\end{figure}

Employing the Hubbard model ($U = 7 |t|$ and $J = 0$) at finite temperature ($k_B T = 0.35 |t|$), we now carry out VWF+$1/\mathcal{N}_f$ analysis of magnetic excitations for fermiology varying in the regime relevant to high-$T_c$ superconductors. Figure~\ref{fig:phasediag_Hubbard} summarizes calculated Hubbard-model paramagnon characteristics as a function of electronic density, $n_e$, on both hole- and electron-doped sides of the phase diagram (dashed vertical lines mark half-filling, $n_e = 1$). Left and right panels correspond to the 3P- and 2P models, respectively. Each curve represents a distinct choice of the range parameter, $r_\mathrm{3P}$ or $r_\mathrm{2P}$, as detailed above the figure. According to the analysis of Sec.~\ref{sec:tb_parametrization}, for high-$T_c$ copper-oxide superconductors  $r_\mathrm{3P} \sim 0.1$-$0.3$ and $r_\mathrm{2P} \sim 0.2$-$0.45$ span distinct (yet overlapping) intervals, which is taken into account in Fig.~\ref{fig:phasediag_Hubbard}. The investigated paramagnon characteristics include propagation energy $\omega_p$ [panels (a) and (e)], bare energy $\omega_0$ [panels (b) and (f)], and damping $\gamma$ [panels (c) and (g)]. Their values have been obtained by secondary DHO modeling of simulated data, following the procedure described in Sec.~\ref{sec:model_method}. Moreover, introducing integrated intensity $I(\omega) \equiv \int_0^\omega d\nu \chi^{\prime\prime}_s(\nu, X)$, in panels (d) and (h) we plot the ratio $I(|t|) / I(\infty)$. This quantity encodes information about relative intensity of the coherent paramagnon peak at low energies (up to the scale of nearest-neighbor hopping integral, $|t|$) and high-energy particle-hole continuum, and may be utilized to probe a crossover between local-moment- and itinerant-electron magnetic dynamics. The dotted line segments near half-filling indicate dynamical instability of the paramagnetic state. The details of the phase stability analysis are presented in Appendix~\ref{appendix:phase_stability}.

Notably, paramagnon propagation frequencies obtained using 3P [Fig.~\ref{fig:phasediag_Hubbard}(a)] and 2P [Fig.~\ref{fig:phasediag_Hubbard}(e)] models exhibit qualitatively distinct dependence on electronic density for large values of range parameter. In particular, the 2P approximation [Fig.~\ref{fig:phasediag_Hubbard}(e)] fails to account for robust paramagnon behavior for $r_\mathrm{2P} \gtrsim 0.4$, since the propagation frequency $\omega_p$ approaches zero on hole-doped side of the phase diagram (cf. black curve). This indicates overdamped magnetic dynamics and is inconsistent with experiment, signaling that the commonly employed 2P Hubbard model \emph{is not} suitable for simultaneous quantitative modeling of fermiology and paramagnon spectra in large-$r_\mathrm{2P}$ high-$T_c$ superconductors, such as multilayer Bi-based cuprates [cf. Fig.~\ref{fig:tb_mapping}(a)]. On the other hand, the microscopically motivated 3P model [Fig.~\ref{fig:phasediag_Hubbard}(a)] yields nonzero values of $\omega_p$ in entire high-$T_c$ range of $r_\mathrm{3P}$, accounting for empirical lack of paramagnon overdamping down to heavily hole overdoped regime. It should be noted though that, for a generic choice $r_\mathrm{2P/3P} = 0.25$, commonly adopted in theoretical work, both parameterizations yield persistent magnetic excitations for $n_e < 1$.

Qualitative differences between the two parameterizations are also apparent on the electron-doped side of the phase diagram. The 2P model yields a systematic hardening of magnetic excitations with electron doping, whereas the 3P model exhibits nonmonotonic behavior of $\omega_p$ (initial hardening is followed by flattening of the $\omega_p$ vs. $n_e$ curve and subsequent decrease of paramagnon propagation frequency). Moreover, the 2P model predicts paramagnon overdamping for $r_\mathrm{2P} \gtrsim 0.4$ in the empirically relevant $1 < n_e \lesssim 1.2$ regime. The overall variation of $\omega_p$ for $1 < n_e < 1.2$ is smaller within the 3P model, and thus is favored by experimental observation of weak dependence of the paramagnon propagation frequency on electron doping in $\mathrm{La_{2-\mathit{x}}Ce_\mathit{x}CuO_4}$ (LCCO) \cite{Li2024}. We note that seemingly conflicting result (paramagnon hardening by $\approx 50 \%$) that would support 2P model has been reported for $\mathrm{Nd_{2-\mathit{x}}Ce_\mathit{x}CuO_4}$ (NCCO) \cite{Lee2014,Ishii2014}. The latter work is, however, based on a simplified Gaussian paramagnon model that does not reflect the DHO propagation frequency $\omega_p$ (cf. the discussion of Ref.~\cite{Li2024}). A transfer of magnetic intensity to higher energies in electron-doped systems is also seen in VWF+$1/\mathcal{N}_f$ simulation results for Hubbard, $t$-$J$-$U$, and $t$-$J$ models \cite{Fidrysiak2023}.

The doping dependence of bare frequency $\omega_0$ and damping $\gamma$ is qualitatively similar for 3P [panels (b)-(c)] and 2P [panels (f)-(g)] models. Both quantities depend weakly on hole concentration, but undergo a systematic enhancement with electron doping (the 2P model on the large-$r_\mathrm{2P}$ end of the high-$T_c$ regime serves as an exception, since enhancement of $\gamma$ close to half-filling results in paramagnon overdamping). The large width of a paramagnon on the electron-doped side of high-$T_c$ phase diagram suggests that magnetic excitations have substantially itinerant character, as also noted previously \cite{Weber2010}. Systematic supression of the ratio $I(|t|)/I(\infty)$ of the low-energy part of the signal to its total integrated magnitude with electron doping provides an independent evidence for a crossover from local-moment to itinerant-electron scenario as a function of electronic density, cf. panels (d) and (h). 

\begin{figure}[!ht]
\centerline{%
\includegraphics[width=12.5cm]{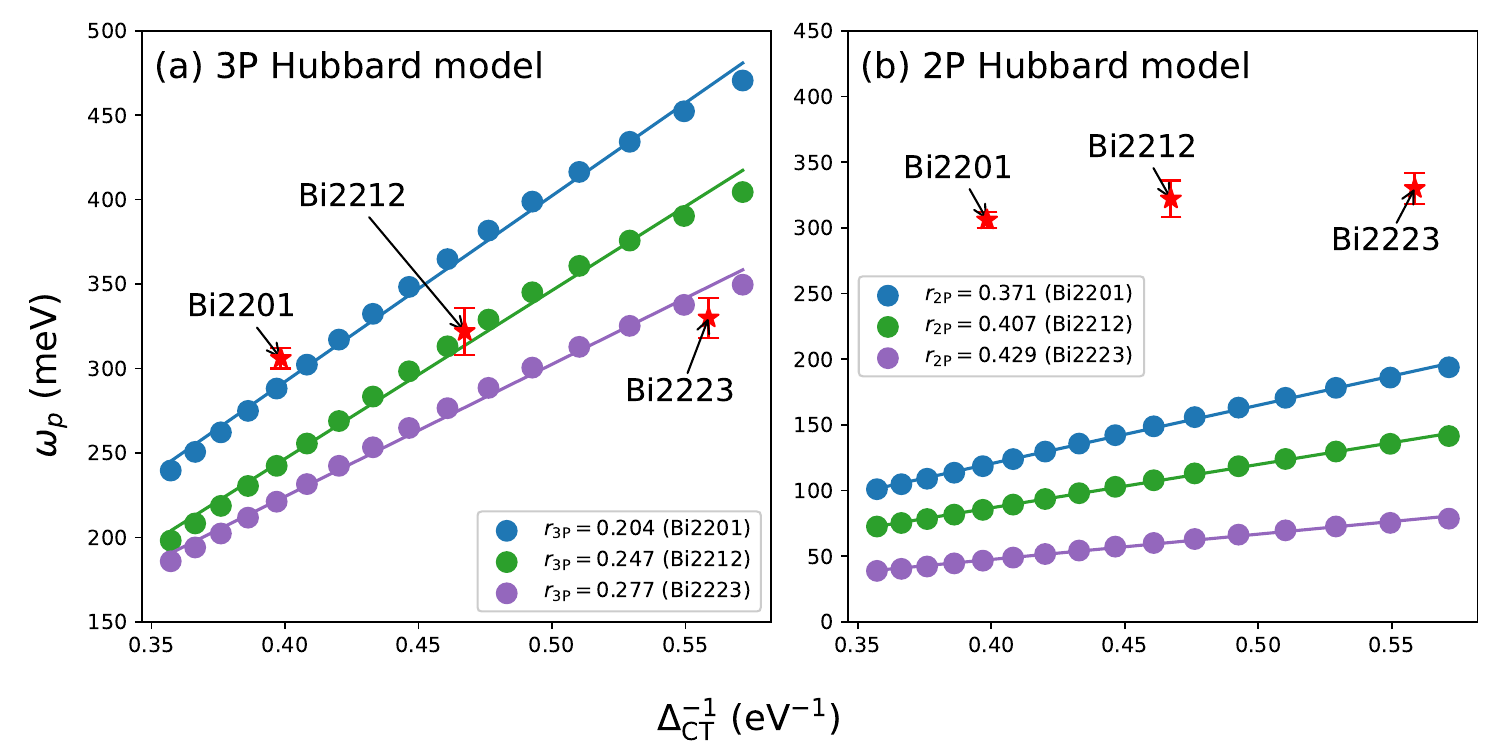}}
\caption{Dependence of the 3P (a) and 2P (b) Hubbard model $X$-point paramagnon propagation energy, $\omega_p$, on inverse charge-transfer gap, $\Delta_\mathrm{CT}^{-1}$. Model parameters are: $n_e = 0.85$, $t = - 0.35\,\mathrm{eV}$, $U = \Delta_\mathrm{CT}$, $J = 0$, and $k_BT = 0.35 |t|$. Range parameters $r_\mathrm{3P}$ ($r_\mathrm{2P}$) are set to experimental values \cite{Lee2006} for Bi-cuprate series, and are detailed inside the panels. Symbols represent VWF+$1/\mathcal{N}_f$ simulation result, and lines show the corresponding linear fits. Red stars are experimental $X$-point propagation energies $\omega^\mathrm{exp}_p \equiv 2 J^\mathrm{exp}$, with effective exchange $J^\mathrm{exp}$ extracted from RIXS data for Bi-family of cuprates \cite{Wang2022}. The 3P-model result [panel (a)] agrees quantitatively with experiment, whereas 2P approximation [panel (b)] fails to account for measured paramagnon energies.}
\label{fig:delta-ct}
\end{figure}

We now demonstrate that the 3P Hubbard model allows for a unified and quantitative interpretation of robust paramagnon physics within Bi-family of cuprates, focusing on representatives hosting $n = 1, 2$, and $3$ $\mathrm{CuO_2}$ planes ($\mathrm{Bi2201}$, $\mathrm{Bi2212}$, and $\mathrm{Bi2223}$, respectively). For those materials, relevant high-quality experimental data have been recently made available, allowing for a stringent test of theory. Figure~\ref{fig:tb_mapping}(a) shows that the ARPES value of the 3P-model range parameter in Bi2223 ($r_\mathrm{3P} \approx 0.28$) is by approximately $40\%$ larger than in Bi2201 ($r_\mathrm{3P} \approx 0.20$). Reading out the corresponding paramagnon energies from Fig.~\ref{fig:phasediag_Hubbard}(a), it is tempting to conclude that the magnetic bandwidth in $\mathrm{Bi2223}$ ($n = 3$) should be by $\sim 50\%$ smaller than in $\mathrm{Bi2201}$ ($n = 1$). This prediction is, however, outright inconsistent with available RIXS measurements \cite{Wang2022} that yield weak (yet systematic) hardening of magnetic excitations with increasing $n$. We show that this discrepancy is resolved once variation of charge transfer energy, $\Delta_\mathrm{CT}$, within the Bi-family of cuprates is taken into account. Empirically, $\Delta_\mathrm{CT}$ decreases from $\approx 2.5\,\mathrm{eV}$ to $\approx 1.85\,\mathrm{eV}$ as $n$ increases from 1 to 3 \cite{Wang2023}. Since high-$T_c$ copper oxides fall into the charge-transfer regime of Zaanen-Sawatsky-Allen classification \cite{Zaanen1985}, their effective on-site Coulomb repulsion $U$ is predominantly governed by $\Delta_\mathrm{CT}$ in the one-band Hubbard model mapping \cite{Feiner1996}. By setting $U \equiv \Delta_\mathrm{CT}$, in Fig.~\ref{fig:delta-ct}(a) we explore theoretically the dependence of the $X$-point paramagnon propagation energy $\omega_p$ on inverse charge transfer energy $\Delta_\mathrm{CT}^{-1}$ for the 3P Hubbard model. Blue-, green-, and purple symbols represent values of $\omega_p$ obtained by VWF+$1/\mathcal{N}_f$ simulation for empirical values of $r_\mathrm{3P}$ corresponding to $\mathrm{Bi2201}$, $\mathrm{Bi2212}$, and $\mathrm{Bi2223}$, respectively [cf. Fig.~\ref{fig:tb_mapping}(a)]. For each considered value of $r_\mathrm{3P}$, paramagnon propagation frequency, $\omega_p$, scales linearly with $\Delta_\mathrm{CT}^{-1}$, which is confirmed by displayed linear fits. In particular, increase of $\Delta_\mathrm{CT}$ results in softening of the anti-nodal paramagnons, similarly as increasing the range parameter $r_\mathrm{3P}$ [see Fig.~\ref{fig:phasediag_Hubbard}(a)]. Given that variations of $r_\mathrm{3P}$ and $\Delta_\mathrm{CT}$ are anticorrelated within Bi-family of cuprates for $n = 1$-$3$, their effects on paramagnon energies are expected to largely cancel each other and result in a moderate dependence of $\omega_p$ on the number of CuO$_2$ layers. Utilizing available data, this conjecture may be tested in a quantitative manner. Corresponding experimental paramagnon propagation frequencies $\omega^\mathrm{exp}_p$ \cite{Wang2022} and charge transfer energies $\Delta^\mathrm{exp}_\mathrm{CT}$ \cite{Wang2023} for consecutive representatives of the Bi-series (hereafter distinguished by a superscript ``exp'') are marked by red symbols in Fig.~\ref{fig:delta-ct}(a). Those empirical data align close to theoretical lines obtained for the respective range parameters $r_\mathrm{3P}$, extracted from an independent experimental probe (ARPES). It should be stressed that, according to our theory, measured weak variation of $\omega^\mathrm{exp}_p$ within Bi-series cannot be explained by considering the effects of variation of either $\Delta_\mathrm{CT}$ or $r_\mathrm{3P}$ alone, but it emerges as a joint effect of single-particle fermiology and electronic correlations. In Fig.~\ref{fig:delta-ct}(b) we have carried out an analysis analogous to that summarized in panel (a), but for the 2P Hubbard model. The employed parameters $r_\mathrm{2P}$ differ from their corresponding $r_\mathrm{3P}$ values as a consequence of nontrivial mapping between 2P and 3P TB Hamiltonians, cf. discussion of Sec.~\ref{sec:tb_parametrization}. Figure~\ref{fig:delta-ct}(b) shows that the 2P model fails to account for experimental data, substantially underestimating measured paramagnon energies. This serves as an independent evidence of 2P model inapplicability to large-$r_\mathrm{2P}$ systems.

\begin{table}
  \centering
  \caption{Comparison of VWF+$1/\mathcal{N}_f$ theory as applied to the 3P Hubbard model with experiment for Bi-family of cuprates (number of $\mathrm{CuO_2}$ layers, $n$, is given next to the cuprate symbol). The empirical quantities reported in the table are: Fermi-surface parameter $r^\mathrm{exp}_\mathrm{3P}$ (ARPES, after~\cite{Lee2006}), charge-transfer energy $\Delta^\mathrm{exp}_\mathrm{CT}$ (STEM-EELS, after~\cite{Wang2023,Wang2023a}), and $X$-point paramagnon propagation energy $\omega^\mathrm{exp}_p \equiv 2 J^\mathrm{exp}$ with $J^\mathrm{exp}$ being measured AF exchange (RIXS, after~\cite{Wang2022} and references therein). The corresponding theoretical VWF+$1/\mathcal{N}_f$ $X$-point paramagnon energies, $\omega^\mathrm{theory}_p$, have been obtained for $n_e = 0.85$, $t = - 0.35\,\mathrm{eV}$, $r_\mathrm{3P} = r^\mathrm{exp}_\mathrm{3P}$, $U = \Delta^\mathrm{exp}_\mathrm{CT}$, and $k_BT = 0.35 |t|$, leaving no more adjustable model parameters. The agreement between $\omega^\mathrm{theory}_p$ and $\omega^\mathrm{exp}_p$ is quantitative. Maximal SC transition temperatures, $T_{c, \textrm{max}}$, are also included for reference (after~\cite{Wang2023a}). \vspace{1em}}
  \label{tab:bi-cuprates}
  \begin{tabular}{ccccccc}
    \hline
    cuprate & $n$ & $T_{c, \textrm{max}}$ (K) & $r^\mathrm{exp}_\mathrm{3P}$ & $\Delta^\mathrm{exp}_\mathrm{CT}$ (eV) & $\omega^\mathrm{exp}_p$ (meV) & $\omega^\mathrm{theory}_p$ (meV) \\
    \hline\hline
    Bi2201 & 1 & 30   &0.204 & 2.51 & 306(6) & 290 \\
    Bi2212 & 2 & 95  & 0.247 & 2.14 & 322(14) & 320 \\
    Bi2223 & 3 & 113  & 0.277 & 1.79 & 330(12) & 343 \\
    \hline
  \end{tabular}
\end{table}

To give a quantitative account of the agreement between theory and experiment, in Table~\ref{tab:bi-cuprates} we compare measured and calculated quantities for Bi-family of cuprates. The number of $\mathrm{CuO_2}$ layers is given next to the compound symbol. The included experimental data are: range parameter $r^\mathrm{exp}_\mathrm{3P}$ (extracted from ARPES \cite{Lee2006}), charge transfer gap $\Delta^\mathrm{exp}_\mathrm{CT}$ (based on outer plane STEM-EELS \cite{Wang2023}), and $X$-point paramagnon energies $\omega^\mathrm{exp}_p \equiv 2 J^\mathrm{exp}$ (with $J^\mathrm{exp}$ obtained from RIXS \cite{Wang2022}). We note that $\omega^\mathrm{exp}_p$ may be also determined directly by analysis of RIXS spectra close to the $X$-point (cf. Supplementary Information of Ref.~\cite{Wang2022}), which results in values consistent with those given in Table~\ref{tab:bi-cuprates} within error bars. The measured quantities, $r^\mathrm{exp}_\mathrm{3P}$ and $\Delta^\mathrm{exp}_\mathrm{CT}$, are used in our microscopic analysis to confine the parameters of the 3P Hubbard model, namely we set $r_\mathrm{3P} \equiv r^\mathrm{exp}_\mathrm{3P}$ and $U \equiv \Delta^\mathrm{exp}_\mathrm{CT}$. The remaining free parameters are electron density $n_e$, nearest-neighbor hopping $t$, and temperature $T$. However, paramagnon propagation energies in the cuprates (both experimentally and theoretically, cf. Fig.~\ref{fig:fit}) are insensitive to $n_e$ across substantial fraction of the hole-doped phase diagram. This circumstance has been instrumental in establishment of the empirical correlation between paramagnon energies and maximal SC transition temperatures \cite{Wang2022}. Without loss of generality, we can thus set density to a representative value close to optimal doping, $n_e = 0.85$, eliminating one more parameter. Moreover, we adopt a generic choice $t = -0.35\,\mathrm{eV}$ \cite{Spalek2022} (the same for all considered representatives of Bi-cuprate series), and set $k_B T = 0.35 |t|$ in order to suppress ordering instabilities. In this way, values of all microscopic parameters are fixed, without further possibility of fine tuning. The resultant theoretical $X$-point paramagnon energies, $\omega_p^\mathrm{theory}$, are reported in the last column of Table~\ref{tab:bi-cuprates}. Their quantitative agreement (maximal deviation $< 6\,\%$) with experimental values $\omega_p^\mathrm{exp}$, extracted independently from RIXS, provides a validation of both 3P Hubbard model and VWF+$1/\mathcal{N}_f$ scheme. We also note that observed weak hardening of magnetic excitations with increasing number of $\mathrm{CuO_2}$ layers is reproduced within our approach. In physical terms, it suggests that paramagnon hardening due to $\Delta_\mathrm{CT}$ reduction overcomes softening effect originating from $r_\mathrm{3P}$ enhancement by a small margin. In Table~\ref{tab:bi-cuprates} we include also maximal SC transition temperatures that undergo enhancement by $\sim 350\,\%$ in the $n = 1$-$3$ interval, and do not scale with paramagnon energy, $\omega_p$. This points toward an indirect relationship between magnetic excitations and SC.

The choice of the same generic value of the hopping integral $t = -0.35\,\mathrm{eV}$ for materials characterized by distinct $\Delta_\mathrm{CT}$ warrants a separate discussion. This is because a qualitative analysis of the three- to one-band model mapping suggests relation $t \propto t_{pd}^2 / \Delta_\mathrm{CT}$, with $t_{pd}$ being $p$-$d$ orbital hopping integral magnitude \cite{Lee2006a}. Assuming formula $J_\mathrm{eff} \propto t^2/\Delta_\mathrm{CT}$, variation of $t$ among the cuprates results thus in scaling $J_\mathrm{eff} \propto t^2/\Delta_\mathrm{CT} \propto\Delta_\mathrm{CT}^{-3}$, rather than $J_\mathrm{eff} \propto \Delta_\mathrm{CT}^{-1}$, seen in Fig.~\ref{fig:delta-ct}(a). Available photoemission data may help to differentiate between those two scenarios by direct mapping of the high-$T_c$ cuprate quasiparticle dispersion. Empirical single-particle spectra separate into three characteristic regimes according to binding energy, i.e. low (below $\approx 10\,\mathrm{meV}$) \cite{Kondo2013}, intermediate (up to the kink at $\approx 70\,\mathrm{meV}$), and high-energy (extending above the kink). The consecutive slopes of electronic dispersion define three velocity scales, $v_\mathrm{low}$, $v_\mathrm{mid}$, and $v_\mathrm{high}$. Whereas $v_\mathrm{low}$ and $v_\mathrm{high}$ vary with hole concentration, $v_\mathrm{mid}$ attains an approximately universal value for multiple cuprates and in broad doping range \cite{Zhou2003}. This universal behavior has been reproduced theoretically, both within the Hubbard \cite{Fidrysiak2018} and $t$-$J$-$U$ \cite{Spalek2017} models, by means of diagrammatic expansion of the Gutzwiller wave function. For Bi-cuprate representatives hosting $n = 1$-$3$ CuO$_2$ planes, laser ARPES yields $1.6\,\text{eV\AA}$ ($n=1$, digitized for optimally doped sample for lowest avaiable temperature) \cite{Peng2013}, $1.8\,\text{eV\AA}$ ($n=2$), and \cite{Vishik2010}  $1.62\,\text{eV\AA}$ ($n=3$) \cite{Chen2025}. Another experiment suggests that both $v_\mathrm{mid}$ and $v_\mathrm{high}$ decease slightly as $n$ increases from $1$ to $2$ \cite{Zhou2003}, albeit their values may be regarded as $n$-independent within error bars. Given that the corresponding $\Delta_\mathrm{CT}$ varies substantially in the $n=1$-$3$ range (cf. Table~\ref{tab:bi-cuprates}), no unambiguous empirical correlation between  $v_\mathrm{mid/high}$ and $\Delta_\mathrm{CT}$ is observed. Assuming that both $v_\mathrm{mid}$ and $v_\mathrm{high}$ are proportional to the effective hopping integral, this points toward a weak dependence of $t$ on $\Delta_\mathrm{CT}$. Microscopic analysis of the three- to one-band model mapping also supports this scenario, revealing breakdown of $t \propto \Delta_\mathrm{CT}^{-1}$ scaling in the charge-transfer regime for intermediate values of oxygen bandwidth and $d$-orbital Coulomb integral $U_d$ \cite{Feiner1996}. Moreover, cellular dynamical mean-field theory \cite{Kowalski2021} calculations suggest weaker dependence of $J_\mathrm{eff}$ on $\Delta_\mathrm{CT}$ than would follow from $J_\mathrm{eff} \sim \Delta_\mathrm{CT}^{-3}$ scaling. From the perspective of the kinematics of paramagnon decays into particle-hole pairs, the relevant range of binding energies is $\sim 0$-$300\,\mathrm{meV}$, leaving the paramagnon exposed to Fermi quasiparticles both below and above the kink. Since the phase space involving exciations governed by $v_\mathrm{low}$ is small, paramagnons are affected predominantly by the intermediate- and high-energy regimes. Assuming in-plane Cu-Cu distance $a = 3.8\,\text{\AA}$, adopted hopping integral $t = -0.35\,\mathrm{eV}$ results in Fermi velocities $v_F = 2.54$, $2.79$, and $2.98\,\text{eV\AA}$ for $n = 1$-$3$ Bi-cuprates, reflecting an averaged value between measured $v_\mathrm{mid} \sim 1.5$-$2\,\text{eV\AA}$ and $v_\mathrm{high} \sim 3$-$6\,\text{eV\AA}$. Parenthetically, obtained by us moderate dependence of $v_F$ on $n$ may be attributed to Gutzwiller band narrowing factor that adjusts to on-site interactions, $U = \Delta_\mathrm{CT}$. Detailed investigation of the effects the kinks in Fermi quasiparticle dispersion on the paramagnon decay kinematics in the cuprates should be the subject of a separate study. 

\section{Paramagnon dynamics: $t$-$J$-$U$ model}
\label{sec:results_tJU_model}

\begin{figure}[!ht]
\centerline{%
\includegraphics[width=12.5cm]{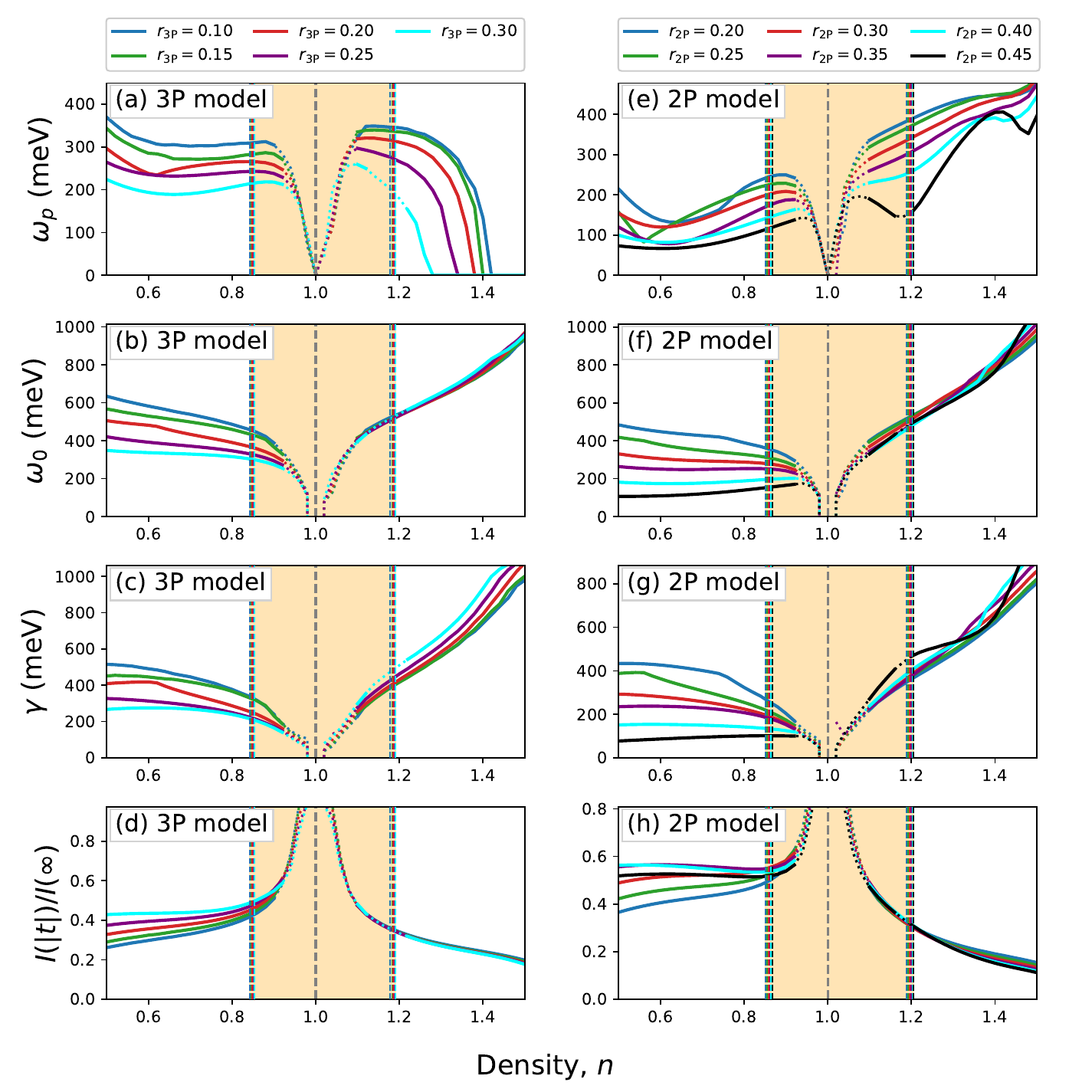}}
\caption{VWF+$1/\mathcal{N}_f$ paramagnon characteristics for the 3P (a)-(d) and 2P (e)-(h) $t$-$J$-$U$ model, to be compared with those of the Hubbard model (cf. Fig.~\ref{fig:phasediag_Hubbard}). Employed range parameters $r_\mathrm{2P/3P}$ are listed above the panels. Dotted line segments indicate dynamical instability of the paramagnetic state, and yellow region marks electronic phase separation. The displayed quantities include: paramagnon propagation energy [(a) and (e)], bare energy [(b) and (f)], and damping [(c) and (g)]. Panels (d) and (h) show the ratio $I(|t|)/I(\infty)$, where $I(\omega) \equiv \int_0^\omega d\nu \chi^{\prime\prime}_s(\nu, X)$ is integrated intensity. The simulations have been carried out for $k_B T = 0.35 |t|$, $t = -0.35\,\mathrm{eV}$, $U = 15 |t|$, and $J = \frac{32}{105} |t| \approx 0.305 |t|$. This results in effective exchange integral $J_\mathrm{eff} = 200\,\mathrm{meV}$.}
\label{fig:phasediag_tJU}
\end{figure}

We now proceed to an analysis analogous to that of Sec.~\ref{sec:results_Hubbard_model}, but employing the $t$-$J$-$U$ model~(\ref{eq:tju-model}) with $U = 15 |t|$ and nonzero $J = \frac{32}{105} |t| \approx 0.305 |t|$. This choice results in effective exchange integral $J_\mathrm{eff} = 200\,\mathrm{meV}$ equal to that considered in Sec.~\ref{sec:results_Hubbard_model}, allowing for a direct comparison between the schemes. The temperature is set to $k_B T = 0.35 |t|$.

Figure~\ref{fig:phasediag_tJU} summarizes relevant paramagnon characteristics for both 3P [panels (a)-(d)] and 2P [panels (e)-(h)] $t$-$J$-$U$ models, arranged in line with Fig.~\ref{fig:phasediag_Hubbard} and following the same notation. A qualitatively new feature, present in Fig.~\ref{fig:phasediag_tJU}, is a broad regime of electronic phase separation around half-filling (yellow area), with boundaries marked by vertical dashed lines. Phase separation is a consequence of strong on-site repulsion ($U = 15|t|$) and high temperature $k_B T = 0.35 |t|$, and is identified based on nonmonotonous dependence of the chemical potential on $n_e$ in the paramagnetic state. The corresponding Maxwell construction is discussed in Appendix~\ref{appendix:phase_stability}.

Figure~\ref{fig:phasediag_tJU}(a) shows that the 3P $t$-$J$-$U$ model yields robust paramagnon behavior on the hole-doped side of the phase diagram. Close to half-filling ($n_e = 1$), paramagnetic state is unstable against fluctuations (cf. dotted segments of the curves) and thus the apparent supression of propagation energy $\omega_p$ for $n_e \rightarrow 1$ is not physically meaningful. We note that the qualitative behavior of $\omega_p$ as a function of $r_\mathrm{3P}$ is the same for the $t$-$J$-$U$ [Fig.~\ref{fig:phasediag_tJU}(a)] and Hubbard [cf. Fig.~\ref{fig:phasediag_Hubbard}(a)] models. With other parameters fixed, increase of $r_\mathrm{3P}$ results in reduction of $\omega_p$, but the quantitative effect is smaller from that observed in the Hubbard model. The 2P $t$-$J$-$U$ model yields substantial softening of paramagnon propagation energy with hole doping (particularly pronounced for generic value $r_\mathrm{2P} = 0.25$), which is inconsistent with experiment. In full analogy with the Hubbard-model analysis, we conclude that the 2P $t$-$J$-$U$ model is not suitable for a joint quantitative analysis of magnetic excitations and fermiology in hole-doped cuprates. Also, the evolution of $\omega_p$ with electron doping is distinct for the 2P and 3P models. Only the 3P-model result may be qualitatively reconciled with a plateau in $\omega_p$ evidenced by recent experiments for electron-doped cuprates \cite{Li2024}. Middle panels of Fig.~\ref{fig:phasediag_tJU} show that both bare paramagnon frequency, $\omega_0$, and damping, $\gamma$, rapidly increase with $n_e$ for $n_e > 1$, pointing toward more incoherent dynamics characteristic of itinerant electrons in electron-doped cuprates. This is confirmed in panels (d) and (h), showing that the ratio $I(|t|)/I(\infty)$ undergoes reduction with increasing electron doping. 

In principle, taking into account the empirical values of range parameter, $r_\mathrm{3P}$, and charge transfer energy, $\Delta_\mathrm{CT}$, one could attempt to carry out a quantitative analysis of paramagnon energies in specific materials, similar to that presented in Sec.~\ref{sec:results_Hubbard_model}. However, at this point there is no unambiguous quantitative mapping of the three-band model of the CuO$_2$ plane onto the one-band $t$-$J$-$U$ model, analogous to that available for the one-band Hubbard model. Incorporation of $\Delta_\mathrm{CT}$ into the one-band $t$-$J$-$U$ Hamiltonian could be achieved, e.g., by imposing the condition  $J_\mathrm{eff} \equiv \frac{4t^2}{\Delta_\mathrm{CT}}$. Given that $J_\mathrm{eff} = \frac{4 t^2}{U} + J$, this results in an expression for explicit AF exchange $J = 4 t^2 \left( \Delta_\mathrm{CT}^{-1} - U^{-1} \right)$. This relation still does not fully determine the parameters of the $t$-$J$-$U$ model, allowing for tuning the on-site repulsion $U$ independently of the value of effective exchange. The detailed properties of this mapping should the analyzed separately.

\section{Summary and outlook}

In this work we have theoretically addressed several aspects relevant to the dynamics of paramagnon excitations in high-$T_c$ cuprates. First, we have constructed a mapping between 2P and 3P tight-binding models of high-$T_c$ fermiology. The simplified two-parameter (2P) approach, commonly used in theoretical work, results in systematically larger parameter $r_\mathrm{2P} \equiv t_\mathrm{2P}^\prime/|t|$ than the corresponding $r_\mathrm{3P} \equiv t_\mathrm{3P}^\prime/|t|$, obtained within the three-parameter (3P) model. This mapping has been verified against available 2P and 3P fits of ARPES Fermi surface for multiple copper oxides, yielding a quantitative agreement.

Subsequently, employing Hubbard model with effective on-site Coulomb repulsion $U = 7 |t|$, we have investigated the impact of high-$T_c$ fermiology (controlled by $r_\mathrm{2P}$ and $r_\mathrm{3P}$ for the 2P and 3P models, respectively) on paramagnon dynamics. Both hole- and electron-doped sides of the phase diagram have been analyzed. Only the 3P-model solution yields robust paramagnon behavior on hole-doped side of the phase diagram in the high-$T_c$ regime, and simultaneously may be qualitatively reconciled with recent experiments on electron-doped cuprates. This signifies the relevance of farther range hopping processes for modeling the high-$T_c$ SC materials. Focusing on the Hubbard model, we have then carried out a quantitative analysis of magnetic excitations for Bi-cuprate family representatives hosting up to three $\mathrm{CuO_2}$ layers. Variational wave function approach, combined with expansion in the inverse number of fermionic flavors, has been employed. Adopting measured values of range parameter $r_\mathrm{3P}^\mathrm{exp}$ and charge-transfer energy $\Delta_\mathrm{CT}^\mathrm{exp}$ (governing the effective Hubbard-$U$ in the one-band model mapping), we have calculated paramagnon propagation energies and demonstrated agreement with experimental values within $6\,\%$ margin. This result establishes a microscopic relationship between data obtained using three experimental probes, in this case two targeting electronic structure (ARPES, STEM-EELS), and one paramagnon dynamics (RIXS). It should be remarked that the relevance of $\Delta_\mathrm{CT}^\mathrm{exp}$ in the one-band Hubbard model analysis of magnetic excitations in high-$T_c$ cuprates, has consequences also to theoretical modeling of a broader class of materials. Among them, infinite-layer (IL) nickelate superconductor Nd$_{1-x}$Sr$_x$NiO$_2$ (NSNO) shares a number of structural and electronic properties with layered copper oxides \cite{Kitatani2020}, but does not exhibit analogous robust paramagnon behavior on the hole-doped side of its phase diagram \cite{Lu2021}. A key difference between IL nickelates and the cuprates is that the former are classified as Mott-Hubbard systems within the ZSA scheme \cite{Goodge2021}. At the level of one-band Hubbard model analysis, this means that the on-site repulsion $U$ is related more directly to correlations within the Cu $d$-electron sector than to the charge-transfer energy, and thus effective $U$ of IL nickelates exceeds that of high-$T_c$ cuprates. Those arguments are sufficient to rationalize relatively small paramagnon bandwidth $\propto \frac{4 t^2}{U}$ in NSNO, but not its qualitatively distinct doping evolution of magnetic excitations. Nonetheless, one-band Hubbard model in the IL-nickelate parameter regime accounts for both bandwidth and doping dependence of paramagnons, pointing toward its applicability to systems on both ends of ZSA classification \cite{Rosa2024}. 

Finally, we have carried out an analysis of the paramagnons within the $t$-$J$-$U$ model, evaluating their characteristics as a function of electronic density and range parameter $r_\mathrm{2P/3P}$. The results are qualitatively consistent with those obtained for the Hubbard model. Yet, an intricate relationship between charge-transfer energy and parameters of the $t$-$J$-$U$ Hamiltonian precludes a direct quantitative analysis that has been possible for the Hubbard model. 

Correlating SC to other material properties has been long considered a route toward identification of key ingredients relevant to high-$T_c$ SC in copper oxides. Maximal SC transition temperatures are known to be linked to the physics of charge transfer between oxygen $2p$ and copper $3d$ orbitals, as evidenced by direct observation of interorbital hole redistribution \cite{Rybicki2016}, as well as indirectly, by measurement of charge transfer energy, $\Delta_\mathrm{CT}$ \cite{Wang2023}. On the other hand, SC also correlates with propagation energies of robust paramagnons, $\omega_p$ \cite{Wang2022}. While empirical relation between $T_c$ and $\Delta_\mathrm{CT}$ emphasizes relationship between local electronic correlations and high-$T_c$ SC, linear scaling of $T_c$ with $\omega_p$ may point toward relevance of paramagnon-driven pairing mechanisms. Our results indicate that there is no one-to-one correspondence between $\omega_p$ and $\Delta_\mathrm{CT}$, providing a microscopic framework for interpretation of such empirical scaling relations. In particular, $n = 1$-$3$ Bi-cuprates exhibit substantial variation of $T_c = 30$-$113\,\mathrm{K}$ \cite{Wang2023a}, despite their nearly identical paramagnon energies (cf. Table~\ref{tab:bi-cuprates}), serving as an exception to the overall linear scaling trend between $T_c$ and $\omega_p$, reported in Ref.~\cite{Wang2022}. The theory of Sec.~\ref{sec:results_Hubbard_model} suggests interpretation of this scaling breakdown in terms of cancellations between the effects of fermiology and electronic correlations on $\omega_p$, even though $\Delta_\mathrm{CT}$ exhibits a pronounced negative correlation with $T_c$. A more detailed investigation of the interplay between electronic properties, magnetic excitations, and high-$T_c$ SC should be undertaken in a separate study.

At the end, we remark on certain plausible extensions of the present analysis. In Sec.~\ref{sec:results_Hubbard_model} we used empirical outer plane (OP) charge-transfer energy. This is well justified for the cuprates with up to $n=3$ planes, where OPs comprise dominant part of bulk system. However, differentiation between OPs and inner CuO$_2$ planes (IPs) allows for a quantitative modeling of $T_c$ evolution as a function of a number of layers \cite{Byczuk1996}, and thus taking it into account is a prerequisite for correlating SC to other microscopic properties. Experiment shows a clear distinction between OPs and IPs for $n \geq 3$ Bi-based cuprates, with $\Delta_\mathrm{CT}$  measured for IP being systematically smaller from the corresponding OP values \cite{Wang2023}. Within the one-band model mapping this implies Hubbard $U_\mathrm{OP} > U_\mathrm{IP}$, where subscripts OP/IP identify the planes. At the same time, high-resolution ARPES for Bi2223 \cite{Chen2025} yields velocity $v_\mathrm{mid}$ nearly the same for IP and OP split bands (supporting comparable effective hopping, $t_\mathrm{IP} \approx t_\mathrm{OP}$), with IP band exhibiting a more pronounced curvature. Moreover, the empirical 3P range parameters of respective CuO$_2$ planes have been suggested to satisfy inequality $r_\mathrm{IP} > r_\mathrm{OP}$ \cite{Ideta2010}. This hierarchy, however, relies on the assumption $t_\mathrm{IP/OP}^{\prime\prime} = - \frac{1}{2} t_\mathrm{IP/OP}^\prime$, and may not hold within more general parameterizations \cite{Luo2023}. Notably, first-principle calculations yield opposite trend \cite{Weber2012}, cf. Fig.~\ref{fig:r_vs_ctgap}. Anticorrelation between $\Delta_\mathrm{OP/IP}$ and $r_\mathrm{OP/IP}$, together with the results of Sec.~\ref{sec:results_Hubbard_model}, would suggest that their contributions to paramagnon propagation energy undergo cancellations not only across materials hosting distinct numbers of CuO$_2$ planes, but also between inequivalent planes of the same compound, resulting in a single layer-independent paramagnon energy scale. A quantitative theoretical verification of this conjecture should, however, involve also hybridization between the planes, which can be effectively incorporated within the variational scheme \cite{Zegrodnik2017}. 

\subsection*{Acknowledgments}

I thank Professor J\'ozef Spa{\l}ek for useful suggestions. This work was supported by Grant Opus UMO-2023/49/B/ST3/03545 from Narodowe Centrum Nauki. For the purpose of Open Access, the author has applied a CC-BY public copyright licence to any Author Accepted Manuscript (AAM) version arising from this submission. The author discloses being one of the Editors of this volume, devoted to the proceedings of Concepts in Strongly Correlated Quantum Matter conference (Kraków, Poland, 2025). 

\subsection*{Data availability}

The dataset containing simulation results presented in this work is available in Ref.~\cite{Fidrysiak2026}.

\appendix

\section{Outline of the VWF+$1/\mathcal{N}_f$ method}
\label{appendix:method}

The VWF+$1/\mathcal{N}_f$ approach extends conventional variational wave function (VWF) method based on minimization of the energy functional

\begin{align}
  E_\mathrm{var} = \frac{\langle \Psi_\mathrm{var}| \hat{\mathcal{H}}| \Psi_\mathrm{var}\rangle}{\langle \Psi_\mathrm{var} | \Psi_\mathrm{var}\rangle}
  \label{eq:Evar}
\end{align}

\noindent
with respect to trial state $|\Psi_\mathrm{var}\rangle \equiv \hat{P}_\mathrm{var} (\boldsymbol{\lambda}) |\Psi_0\rangle$. Here $\hat{P}_\mathrm{var}(\boldsymbol{\lambda})$ is an operator introducing correlations intro Slater determinant state $|\Psi_0\rangle$, dependent on a vector of parameters $\boldsymbol{\lambda}$ to be determined in the optimization procedure. For compactness of notation, we assume that the chemical potential, $\mu$, has been incorporated into the Hamiltonian, i.e. $\hat{\mathcal{H}} \leftrightarrow \hat{\mathcal{H}} - \mu \hat{N}$ with $\hat{N}$ being particle number operator. Following application of Wick's theorem, variational energy $E_\mathrm{var} = E_\mathrm{var}(\mathbf{P}, \boldsymbol{\lambda}, \mu)$ may be expressed as functional of two-point correlation functions $P_{ij\sigma\sigma^\prime} \equiv \langle \hat{a}_{i\sigma}^\dagger \hat{a}_{j\sigma^\prime} \rangle_0$  (collectively denoted as $\mathbf{P}$ and referred to as ``lines''), correlator parameters $\boldsymbol{\lambda}$, and chemical potential $\mu$. Here $\langle \ldots \rangle_0$ represents uncorrelated expectation value, evaluated with the Slater determinant state, $| \Psi_0 \rangle$. It is also useful to introduce an analogous notation for Fermi bilinears $\hat{P}_{ij\sigma\sigma^\prime} \equiv \hat{a}_{i\sigma}^\dagger \hat{a}_{j\sigma^\prime}$ (comprising a vector $\hat{\mathbf{P}}$). This allows us to write down a compact identity  $\mathbf{P} = \langle\hat{\mathbf{P}}\rangle_0$. To improve the convergence properties of the diagrammatic expansion of $E_\mathrm{var}$, multiple additional constraints for the variational parameters need to be imposed \cite{Buenemann2012,Kaczmarczyk2014}. Here we denote them symbolically as $\mathbf{C}(\mathbf{P}, \boldsymbol{\lambda}) = \mathbf{0}$; for a derivation of their form suitable for the analysis of paramagnon dynamics, see Refs.~\cite{Fidrysiak2021,Spalek2022}.

Instead of carrying out a direct constrained optimization of $E_\mathrm{var}$, we define the Landau functional
\begin{align}
  \mathcal{F}(\mathbf{P}, \boldsymbol{\lambda}, \boldsymbol{\xi}, \boldsymbol{\rho}, \mu) = -\frac{1}{\beta} \ln \mathrm{Tr} \exp \left( - \beta \hat{\mathcal{H}}_\mathrm{eff} \right),
  \label{eq:Landau_fucntional}
\end{align}

\noindent 
where
\begin{align}
  \hat{\mathcal{H}}_\mathrm{eff}(\mathbf{P}, \boldsymbol{\lambda}, \boldsymbol{\xi}, \boldsymbol{\rho}, \mu) = E_\mathrm{var}(\mathbf{P}, \boldsymbol{\lambda}, \mu) - i \boldsymbol{\xi}^\dagger \left(\mathbf{P} - \hat{\mathbf{P}}\right) - i \boldsymbol{\rho}^T \mathbf{C}(\mathbf{P}, \boldsymbol{\lambda})
  \label{eq:effective_Hamiltonian}
\end{align}

\noindent
is the effective Hamiltonian and $\beta \equiv (k_B T)^{-1}$. Variables $\boldsymbol{\xi}$ and $\boldsymbol{\rho}$ serve as Lagrange multipliers ensuring that $\mathbf{P} \equiv \langle \hat{\mathbf{P}} \rangle$ and $\mathbf{C}(\mathbf{P}, \boldsymbol{\lambda}) \equiv 0$, respectively. Here $\langle \ldots \rangle$ refers to thermal expectation value. The components of $\boldsymbol{\xi}$ are of the form $\xi_{ij\sigma\sigma^\prime}$, reflecting the structure of lines $P_{ij\sigma\sigma^\prime}$. Note that, by construction, $\hat{P}^\dagger_{ij\sigma\sigma^\prime} = \hat{P}_{ji\sigma^\prime\sigma}$. In our analysis we impose the same symmetry relation on $\mathbf{P}$ and $\boldsymbol{\xi}$, i.e. $P^{*}_{ij\sigma\sigma^\prime} =P_{ji\sigma^\prime\sigma}$ and $\xi^{*}_{ij\sigma\sigma^\prime} = \xi_{ji\sigma^\prime\sigma}$. In particular, the diagonal lines, $P_{ii\sigma\sigma}$, representing local electronic density, are real. The formulation of Eq.~(\ref{eq:Landau_fucntional}) has been originally introduced for Gutzwiller-type wave function, resulting in the so-called statistically-consistent Gutzwiller approximation (SGA) \cite{Jedrak2010}; the term \emph{statistical consistency} refers to the condition $\mathbf{P} \equiv \langle \hat{\mathbf{P}} \rangle$ enforced by Lagrange multiplier method. In the zero-temperature limit ($\beta \rightarrow \infty$), the saddle point of the functional~(\ref{eq:Landau_fucntional}) with respect to variables $\mathbf{P}$, $\boldsymbol{\lambda}$, $\boldsymbol{\xi}$, and $\boldsymbol{\rho}$ provides a necessary condition for a constrained minimum of Eq.~(\ref{eq:Evar}) at fixed $\mu$. Chemical potential term is determined by solving the condition for electron number $N_e = - \frac{\partial \mathcal{F}}{\partial \mu}$. Notably, the formulation of Eq.~(\ref{eq:Landau_fucntional}) is applicable also at finite temperature, which we utilize in the discussion of paramagnon dynamics.

The VWF+$1/\mathcal{N}_f$ approach is based on Eqs.~(\ref{eq:Evar}) and (\ref{eq:effective_Hamiltonian}), generalized to a dynamical situation. Specifically, the fields are promoted to imaginary-time-dependent quantities and decomposed as $\mathbf{P}(\tau) = \mathbf{P}_0 + \delta \mathbf{P}(\tau)$, $\boldsymbol{\lambda}(\tau) = \boldsymbol{\lambda}_0 + \delta \boldsymbol{\lambda}(\tau)$, $\boldsymbol{\xi}(\tau) = \boldsymbol{\xi}_0 + \delta \boldsymbol{\xi}(\tau)$, and $\boldsymbol{\rho}(\tau) = \boldsymbol{\rho}_0 + \delta \boldsymbol{\rho}(\tau)$, where subscript ``0'' indicates the saddle-point value. Hamiltonian is then substituted with the action

\begin{align}
  \mathcal{S} = & \sum_{\substack{ij\sigma\sigma^\prime \\ s=1\ldots \mathcal{N}_f}} \int_0^\beta  d\tau \bar{\eta}^s_{i\sigma} \left( \partial_\tau \delta_{ij} \delta_{\sigma\sigma^\prime} + i \xi^{*}_{ij\sigma\sigma^\prime}  \right)\eta^s_{j\sigma^\prime} + \nonumber \\ & \mathcal{N}_f \int_0^\beta d\tau \left(E_\mathrm{var}(\mathbf{P}, \boldsymbol{\lambda}, \mu) - i \boldsymbol{\xi}^\dagger \mathbf{P} - i \boldsymbol{\rho}^T \mathbf{C}(\mathbf{P}, \boldsymbol{\lambda}) \right) + \text{$\mathcal{O}(1)$ terms},
\label{eq:action}
\end{align}

\noindent
where $\eta_{i\sigma}^s$ and $\bar{\eta}_{i\sigma}^s$ represent $s = 1, \ldots, \mathcal{N}_f$ families of Grassman fields, governing itinerant electron dynamics. The grand potential then reads $\Omega = -\frac{1}{\beta} \ln Z$, with $Z = \int \exp(-\mathcal{S})$. The $\mathcal{O}(1)$ terms, as well as several technical contributions not explicitly included in $\mathcal{S}$, are needed to rationalize transition from Eq.~(\ref{eq:effective_Hamiltonian}) to Eq.~(\ref{eq:action}), and for convergence purposes \cite{Fidrysiak2021}.

Physically, there is only one set of fermionic fields ($\mathcal{N}_f = 1$), but $1/\mathcal{N}_f$ may be regarded as a formal parameter controlling perturbation expansion. We work in the $\mathcal{N}_f \rightarrow \infty$ limit which allows us to derive a closed form expression for the  dynamical susceptibility matrix $\hat{\chi} = (1 + \hat{\chi}_0 \hat{\mathcal{V}}_\mathrm{eff})^{-1} \hat{\chi}_0$. Here $\hat{\chi}_0$ is the Lindhard susceptibility, calculated using the effective Hamiltonian~(\ref{eq:effective_Hamiltonian}). As a result, already the fermion loop integral $\hat{\chi}_0$ incorporates certain correlation effects, such as bandwidth normalization. The other quantity, contributing to $\hat{\chi}$, is the matrix $\hat{\mathcal{V}}_\mathrm{eff}$, representing effective interactions between Landau quasiparticles. The formal structure of $\hat{\chi}$ is reminiscent of that resulting from random phase approximation. However, it should be emphasized that $\hat{\mathcal{V}}_\mathrm{eff}$ contains not only renormalized interactions already present in the original Hamiltonian $\hat{\mathcal{H}}$, but also qualitatively new nonlocal matrix elements. This is a dynamical manifestation of the circumstance that local interactions induce nonlocal correlations, particularly close to metal-insulator transition.

Following a general overview of VWF+$1/\mathcal{N}_f$ approach, we now specify relevant details of the adopted variational wave function. We consider $\hat{P}_\mathrm{var} = \prod_i \hat{P}_{\mathrm{var}, i}$ in a product form, where

\begin{align}
  \hat{P}_{\mathrm{var}, i} \equiv \lambda^0_{i} |0\rangle_{ii}\langle 0| + \sum_{\sigma \sigma^\prime} \lambda^{\sigma\sigma^\prime}_{i}|\sigma\rangle_{ii}\langle\sigma^\prime| + \lambda^d_{i} |{\uparrow\downarrow}\rangle_{ii}\langle {\uparrow\downarrow}|
  \label{eq:local_correlator}
\end{align}

\noindent
represents local correlator, controlled by six coefficients $\lambda^0_{i}$, $\lambda^{\sigma\sigma^\prime}_{i}$ (with $\sigma, \sigma^\prime = \uparrow, \downarrow$), and $\lambda^d_{i}$. Their values are restricted by the condition $\hat{P}_{\mathrm{var}, i}^\dagger = \hat{P}_{\mathrm{var}, i}$, as well as by the constraints $\mathbf{C}(\mathbf{P}, \boldsymbol{\lambda}) \equiv \mathbf{0}$. The states $|0\rangle_{i}$, $|{\uparrow}\rangle_{i}$, $|{\downarrow}\rangle_{i}$, and $|{\uparrow\downarrow}\rangle_{i}$ form the local basis on lattice site $i$. The structure of the correlator~(\ref{eq:local_correlator}) allows for reweighing local many-body configurations in response to interactions, according to the variational procedure. We note that the off-diagonal coefficients, $\lambda^{\uparrow\downarrow}_{i}$ and $\lambda^{\downarrow\uparrow}_{i}$, are usually neglected in equilibrium paramagnetic-state simulations. The role of those terms in the present study is to ensure that the variational wave function is general enough to accommodate noncollinear spin configurations associated with paramagnon excitations. The energy functional~(\ref{eq:Evar}) is evaluated to the leading order in the systematic real-space diagrammatic expansion \cite{Buenemann2012,Kaczmarczyk2014}, which is equivalent to the SGA expression. 

\begin{figure}[t]
\centerline{%
\includegraphics[width=12.5cm]{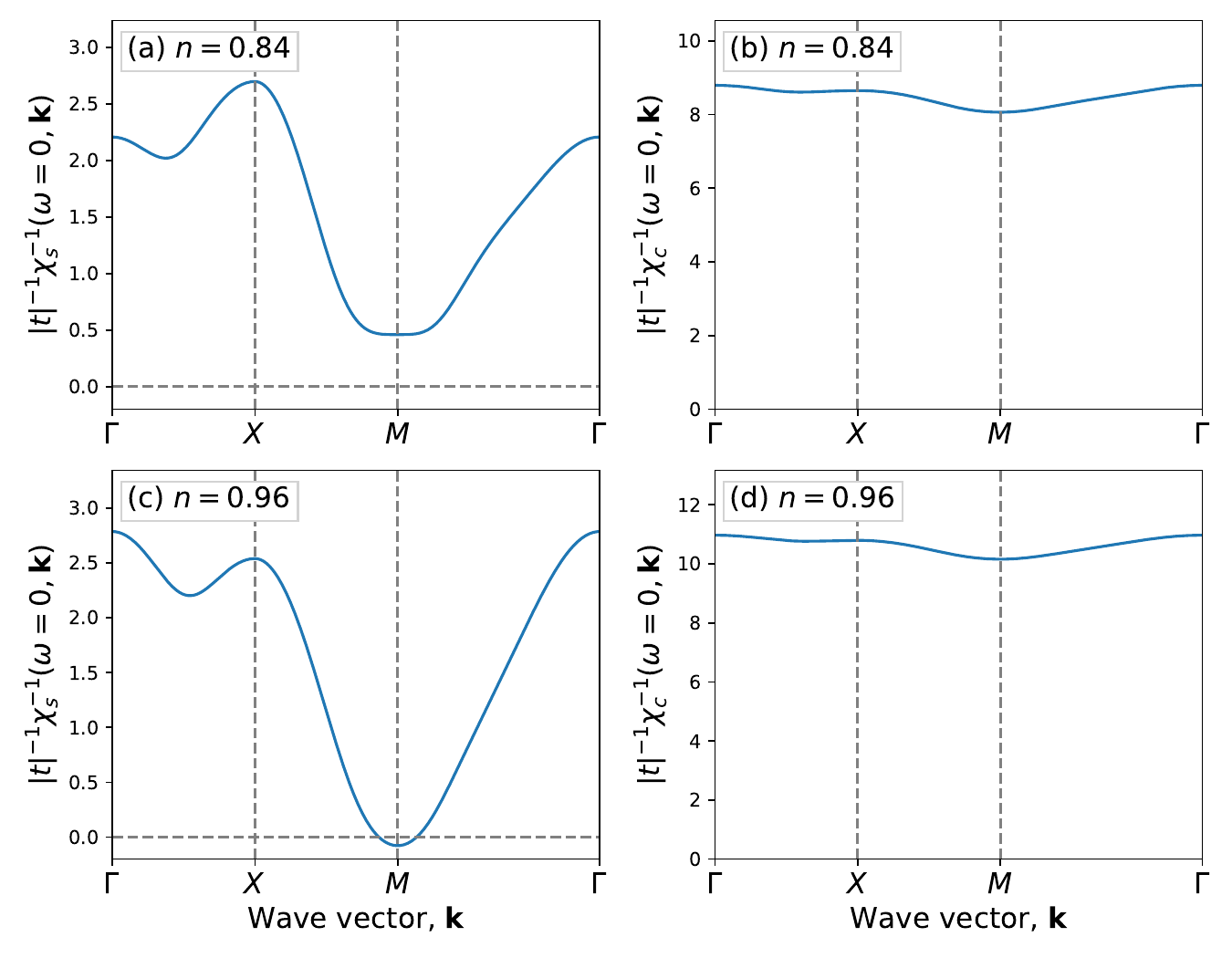}}
\caption{Analysis of the paramagnetic state stability against fluctuations for the 3P Hubbard model ($U = 7 |t|$, $J = 0$, $r_\mathrm{3P} = 0.20$, $k_B T = 0.35 |t|$), and representative densities, $n_e = 0.84$ [(a)-(b)] and  $n_e = 0.96$ [(c)-(d)]. Left (right) panels show the inverse of static spin (charge) susceptibility along the high-symmetry $\Gamma$-$X$-$M$-$\Gamma$ Brillouin-zone contour, denoted as $\chi_s^{-1}(\omega = 0, \mathbf{k})$ and $\chi_c^{-1}(\omega = 0, \mathbf{k})$, respectively. Panel (c) indicates antiferromagnetic instability.} 
\label{fig:phase_stab}
\end{figure}

\section{Phase stability analysis}
\label{appendix:phase_stability}

Here we summarize the analysis of paramagnetic state stability. Left (right) panels of Fig.~\ref{fig:phase_stab} show inverse of static spin (charge) susceptibilities along the high-symmetry $\Gamma$-$X$-$M$-$\Gamma$ Brillouin-zone contour, calculated for the 3P Hubbard model. The employed parameters, $U = 7 |t|$, $J = 0$, $r_\mathrm{3P} = 0.20$, and $k_B T = 0.35 |t|$, are representative of those used in the main text. For electronic density $n_e = 0.84$ [panels (a)-(b)], both spin- and charge susceptibilities remain positive, reflecting stability of the paramagnetic state against fluctuations. This is not the case for $n_e = 0.96$ close to half filling [panels (c)-(d)], where spin susceptibility becomes negative close to the $M$ point. This indicates instability toward commensurate antiferromagnetic order. In Figs.~\ref{fig:phasediag_Hubbard} and \ref{fig:phasediag_tJU}, dotted lines represent solutions that are unstable according to this procedure. We have also verified that there are no local instabilities along $\Gamma$-$X$-$M$-$\Gamma$ contour in the parameter range used to compose Fig.~\ref{fig:delta-ct} of the main text.

\begin{figure}[t]
\centerline{%
\includegraphics[width=12.5cm]{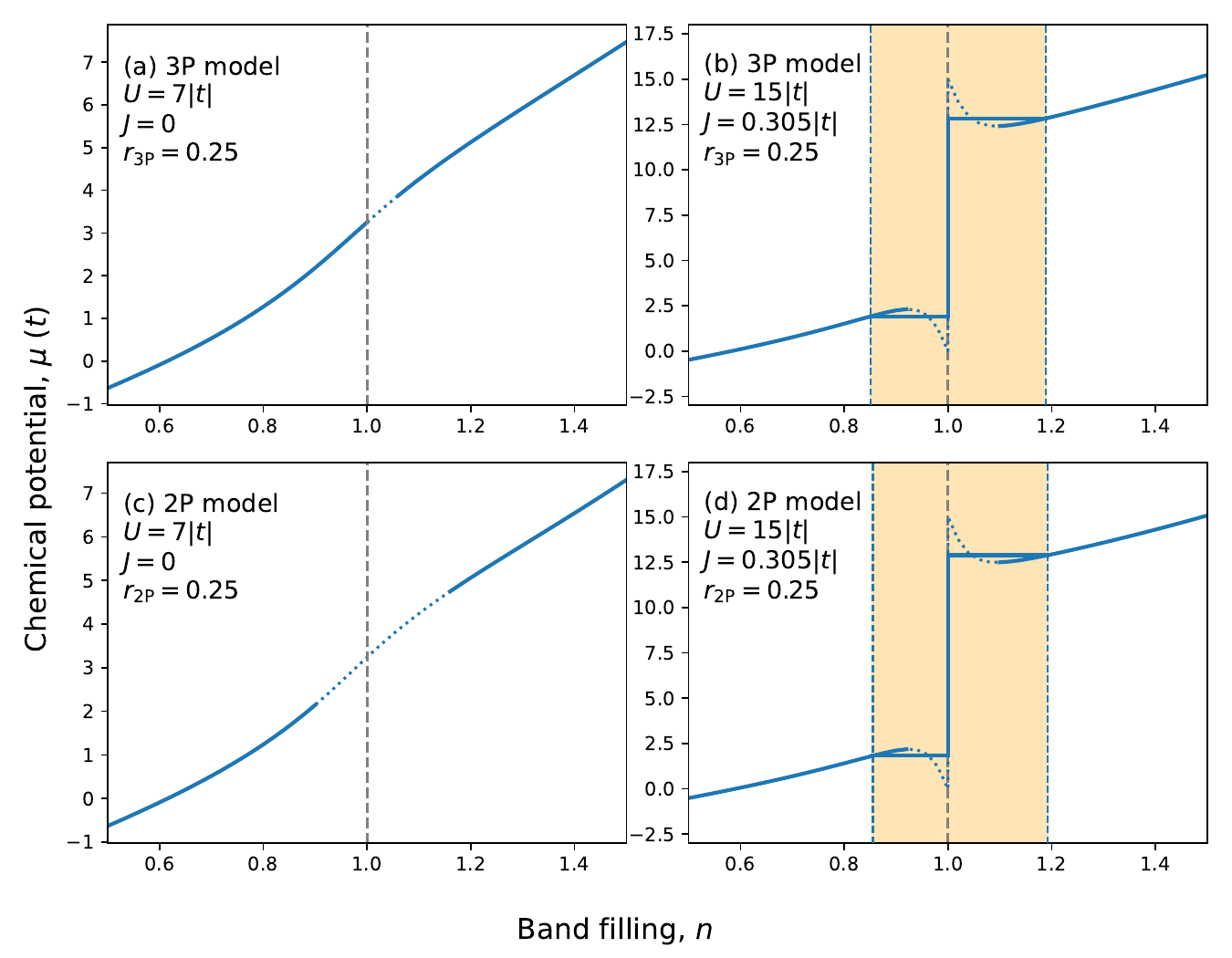}}
\caption{Chemical potential $\mu$ (solid lines) as a function of electronic density $n$ for (a) 3P Hubbard model, (b) 3P $t$-$J$-$U$ model, (c) 2P Hubbard model, and (d) 2P $t$-$J$-$U$ model. The temperature is set to $k_B T = 0.35 |t|$, and the remaining parameters are listed inside the panels. Dotted line segments indicate local instabilities against fluctuations, obtained by the procedure depicted in Fig.~\ref{fig:phase_stab}. For both 2P and 3P $t$-$J$-$U$ model, $\mu$ is a nonmonotonic function of $n$, which signals phase separation. Yellow area in panels (b) and (d) marks phase separation region as obtained using Maxwell construction.}
\label{fig:chemical_potential}
\end{figure}

The phase stability analysis for the $t$-$J$-$U$ model requires a separate discussion. This is because, in addition to local instabilities against fluctuations, paramagnetic metallic state is prone to phase separation due large on-site repulsion $U$ and elevated temperature. Figure~\ref{fig:chemical_potential} shows doping dependence of chemical potential, $\mu$, plotted as a function of density, $n$, for (a) 3P Hubbard-, (b) 3P $t$-$J$-$U$-, (c), 2P Hubbard-, and (d) 2P $t$-$J$-$U$ models. Temperature has been set to $k_B T = 0.35 |t|$, and the remaining parameters are detailed inside the panels. The dotted lines mark local fluctuation-driven instabilities, obtained using on the procedure illustrated in Fig.~\ref{fig:phase_stab}. For the Hubbard model [panels (a) and (c)] $\mu$ is an increasing function of density, whereas chemical potential exhibits a nonmonotonic behavior for the $t$-$J$-$U$ model. Negative slope of the function $\mu(n)$ indicates negative compressibility, signaling electronic phase separation [yellow regions in panels (b) and (d)]. The phase separation boundaries are determined quantitatively using Maxwell construction, as illustrated inside the figure. We note that the extent of local fluctuation-driven instabilities (dotted line segments) for the $t$-$J$-$U$ model does not coincide with the phase separation regime. This points toward emergence of metastable solutions within the first-order phase transition region.

In our analysis, we have determined the boundaries of phase separation by inspecting paramagnetic-state behavior of $\mu(n)$ function. For most of the parameter configurations considered in the main text this is justified, since phase separation occurs before any local fluctuation-driven instabilities along the $\Gamma$-$X$-$M$-$\Gamma$ high-symmetry contour are observed. However, for the 3P $t$-$J$-$U$ model with $r_\mathrm{3P} = 0.30$ (cf. Fig.~\ref{fig:phasediag_tJU}), local instability on the electron-doped side precedes the phase separation. A more detailed analysis should then be based on the analysis of the chemical potential calculated in the resultant broken-symmetry state, with appropriately readjusted phase separation boundary. 

\printbibliography

\end{document}